\documentclass[epj]{svjour}
\usepackage{graphics}
\begin{document}
\title{A new design for the CERN-Fr\'ejus neutrino Super Beam}
\author{A. Longhin}
\institute{ 
 Irfu, CEA-Saclay, 91191 Gif-sur-Yvette, France
\thanks{\emph{Now at Laboratori Nazionali  di Frascati, INFN, Italy}}
}
\date{
}
\abstract{
We present an optimization of the hadron focusing system for a low-energy 
high intensity conventional neutrino beam (Super-Beam)
proposed on the basis of the
HP-SPL at CERN with a beam power of 4 MW and an energy of 4.5~GeV. The
far detector would be a 440~kton Water Cherenkov detector (MEMPHYS)
located at a baseline of 130~km in the Fr\'ejus site.
The neutrino fluxes simulation relies on a new GEANT4 based simulation coupled with an optimization
algorithm based on the maximization of the sensitivity limit on the $\theta_{13}$
mixing angle. 
A new configuration adopting a multiple horn system with solid targets
is proposed which improves the sensitivity to
$\theta_{13}$ and the CP violating phase $\delta_{CP}$.
}
\maketitle
\section{Physics goals}
\label{intro}
The determination of the $\nu_\mu \to \nu_e$ oscillation probability 
is one of the main goals of current research in neutrino physics. 
This process contains information on currently unknown fundamental
parameters of the PMNS matrix which describes the mixing between the neutrino mass and flavor eigenstates:
the $\theta_{13}$ angle and the $\delta_{CP}$ phase. Moreover the ordering scheme of neutrino
masses (the so called mass hierarchy) could be determined through effects induced by neutrino interactions
with matter through the MSW effect.
Given a non null $\theta_{13}$, CP violating effects generated by $\delta_{CP}\neq 0$ or $\pi$ would 
induce a difference in the oscillating behavior of neutrinos and anti-neutrinos: 
$\mathcal{P}(\nu_\mu\to\nu_e)$ $\neq$ $\mathcal{P}(\bar{\nu}_\mu\to\bar{\nu}_e)$.
The presence of such effects involving light neutrinos would 
support the leptogenesis hypothesis as an explanation for the
observed matter/anti-matter asymmetry in the Universe.

With respect to $\nu_\mu\to\nu_\tau$ transitions for which the mixing is maximal,
$\nu_e$ appearance is a sub-leading effect suppressed by the smallness of the $\theta_{13}$
angle. This imposes the use of a very intense and pure neutrino beam and a Mton scale low density detector
capable of measuring the appearance of $\nu_e$. 

In this work we present a development of the original proposal for a high-intensity
conventional neutrino beam (Super-Beam) produced at CERN and aimed towards a 440~kton
Water Cherenkov detector at the LSM laboratories in the Fr\'ejus tunnel, at a distance of 130~km. 

This relatively short baseline is beneficial in terms of suppression of the neutrino flux. Furthermore 
the sub-GeV spectrum is well matched to the region in which quasi-elastic interactions
are dominant. Water Cherenkov detectors provide excellent reconstruction for this
topology with good rejection capability for the $\pi^0$ mesons produced in neutral current interactions.
Furthermore a clean measurement of CP violation (CPV) is possible since there is almost no interplay with effects 
related to propagation in matter. 

First studies \cite{Ball:2000pj,Gilardoni:2004kr,Mezz03}
were assuming a 2.2~GeV proton beam and a
liquid mercury jet target associated with a 
single conic horn with a pulsed current of 300~kA. 
Later it was proposed \cite{Gilardoni:2003us} to supplement the system with an auxiliary horn 
(called reflector) enclosing concentrically the first horn and operated at 600~kA
 in order to focus also pions produced at larger angles. 
This scheme was adopted in \cite{Campagne:2004wt} and the horn shape re-optimized 
using the method described in  	\cite{Campagne:2004cd}. Further, the decay tunnel was re-optimized using
different primary beam energies from 2.2 up to 8~GeV. Based on the neutrino fluxes of \cite{Campagne:2004wt} 
and an improved parametrization of the far detector, the physics performances of the project were presented 
in \cite{Campagne:2006yx} assuming a 3.5~GeV proton kinetic energy.

With respect to previous studies on this subject we propose a
new design based on the use of a solid target and a single magnetic horn 
operated with a lower value of the pulsed current (300-350~kA).
Such a setup simplifies the engineering complexity of the system 
avoiding difficult issues such as the containment of the mercury jet in a magnetic field free region \cite{cit:MERIT},
the challenge of a power supply operating at 600~kA and 
the constraints related to mechanical stresses on the horn-reflector system induced by the high frequency 
current pulsing. 
The capability for a solid target to sustain for a reasonable
time a power of 4 MW has not yet been demonstrated. For this reason we propose a setup based on a
battery of four target-horn systems operated sequentially \cite{Gilardoni:2001gv}.

In Section \ref{sect:setup} we describe the proton source and far detector (\ref{sect:pd}),  
the target (\ref{sect:target}) and horn (\ref{sect:horn}) setup, the method used to calculate the neutrino 
fluxes (\ref{sect:fluxescalc}) and the 
sensitivity to the physical parameters (\ref{sect:limits}). 
Section \ref{sect:OPT} describes a systematic procedure 
of optimization of the horn and decay tunnel based on the sensitivity on $\sin^22\theta_{13}$.
The neutrino fluxes obtained with an optimized setup are the content of Section \ref{sect:resOPT} together with the sensitivities to 
$\sin^22\theta_{13}$ and $\delta_{CP}$ for which we also
take into account the uncertainty related to hadroproduction correcting the simulation with experimental data. 
Finally Section \ref{sect:alt} describes two alternative options for the target design.

\section{Setup}
\label{sect:setup}
\subsection{Proton driver and far detector}
\label{sect:pd}
The proton driver we consider is the high power version of the Super
Conducting Proton Linac (HP-SPL) at CERN. 
The current design study
\cite{CERN:SPL:design} 
foresees a proton kinetic energy of 4.5~GeV and a beam power of $4$~MW at $50$~Hz repetition
frequency with a pulse duration of about 400~$\mu$s.
An accumulator ring divides the protons into pulses of $\simeq 1\,\mu$s 
with a sub-structure of either 6, 3 or 1 bunches per pulse \cite{NuFact:acc-cmp:ring}.
It should be noted that this machine 
supplemented by a compressor ring could constitute the 
proton driver for
a future 
neutrino-factory.

The far detector which we consider is a water-Cherenkov with a 440~kton fiducial mass 
(MEMPHYS) \cite{deBellefon:2006vq}. A design with a mass of 500~kton is
also being considered. Besides its role in neutrino oscillation physics, the programme 
of such a detector includes the study of proton decay, atmospheric and SuperNovae neutrinos.

\subsection{Target}
\label{sect:target}

Due to the low energy of the proton beam pions are emitted at relatively large angles.
This constraint forces to place the target inside the horn in order to preserve a good 
collection efficiency.

In previous studies the liquid mercury 
target has been schematically modeled as a cylinder with 0.75~cm radius and 30~cm
length corresponding to about two hadronic interaction lengths
($\lambda_I$). In this study graphite was chosen as an alternative material since
solid targets composed of graphite constitute a proven technology in existing neutrino beams (i.e. T2K and CNGS).
It must however be noted that other low atomic number (low-$Z$) materials such as Beryllium, Aluminum or 
AlBeMet\textregistered~(38\% Al, 62\% Be) give similar
particle multiplicities and spectra. 
A granular Titanium structure has also been considered and is discussed in Sect. \ref{sect:alt}.

\begin{figure}
\centering
\resizebox{0.5\textwidth}{!}{
\includegraphics{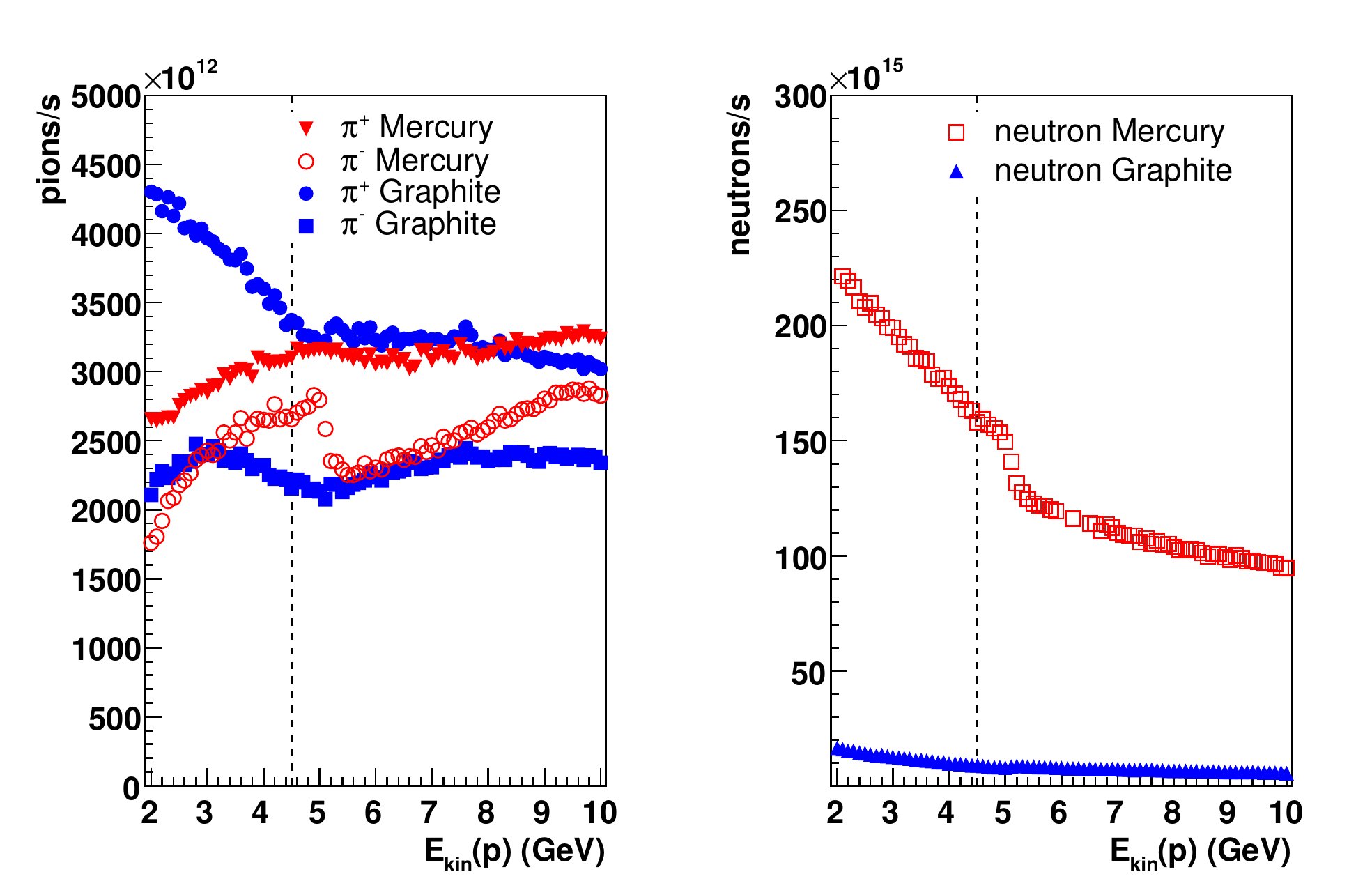}
}
\caption{Particle yields versus the proton kinetic energy for a 30~cm long, 1.5~cm diameter mercury target and a 78~cm long 1.5~cm diameter graphite target 
with a 4 MW beam power according to the FLUKA simulation. 
}
\label{fig:yieldsC}
\end{figure}

In \cite{NUFACT09} a comparison of the mercury and the carbon 
target in terms of energy deposition and secondary particle yields is drawn.
The power deposited in the target at 4.5~GeV and 4~MW power is about 220~kW
and 700~kW for graphite and mercury respectively. 
Pion yields are similar while the most remarkable difference between the two targets is the
neutron yield which is reduced by a factor 15 for the graphite target (see Fig.~\ref{fig:yieldsC}).  
This reduction is beneficial in view of the radiation damage caused by neutrons on 
the horn\cite{ncampagne}.
In this work we consider a cylindrical target with a radius of 1.5~cm, a length of 78~cm
and a density of 1.85 g/cm${}^3$ corresponding to roughly 2 $\lambda_I$.

\subsection{Horn
\label{sect:horn}}

In the previous study the horn shape is 
designed 
in order to focus 
pions of $\simeq$ 600 MeV to produce neutrinos with energies at the oscillation maximum
($E_\nu \simeq$ 260 MeV).  Assuming a point-like source and a
cylindrical surface close to the target the optimal shape is roughly
conical. 

In \cite{NUFACT09} we have
shown that the conical horn optimized for the liquid mercury target is not suited to
a long solid target since the $\sin^2 2\theta_{13}$ sensitivity limit 
for $\delta_{CP}<\pi$ degrades significantly. The effect was found to be caused by
the contamination of $\nu_e$ in the $\bar{\nu}_\mu$ beam from cascade decays
of positive pions ($\pi^+\to \mu^+\to e^+ \nu_e \bar{\nu}_\mu$) and kaons ($K^+\to \pi^0e^+\nu_e$) 
which are not effectively defocused when they are produced in the forward direction\footnote{The equivalent 
effect is weaker in the $\nu$ beam due to
the combined effect of the reduced cross section of $\bar{\nu}_e$ and the fact
that negatively charged hadrons are less abundantly produced.}.  

\begin{figure}
\begin{center}
\resizebox{0.5\textwidth}{!}{
\includegraphics{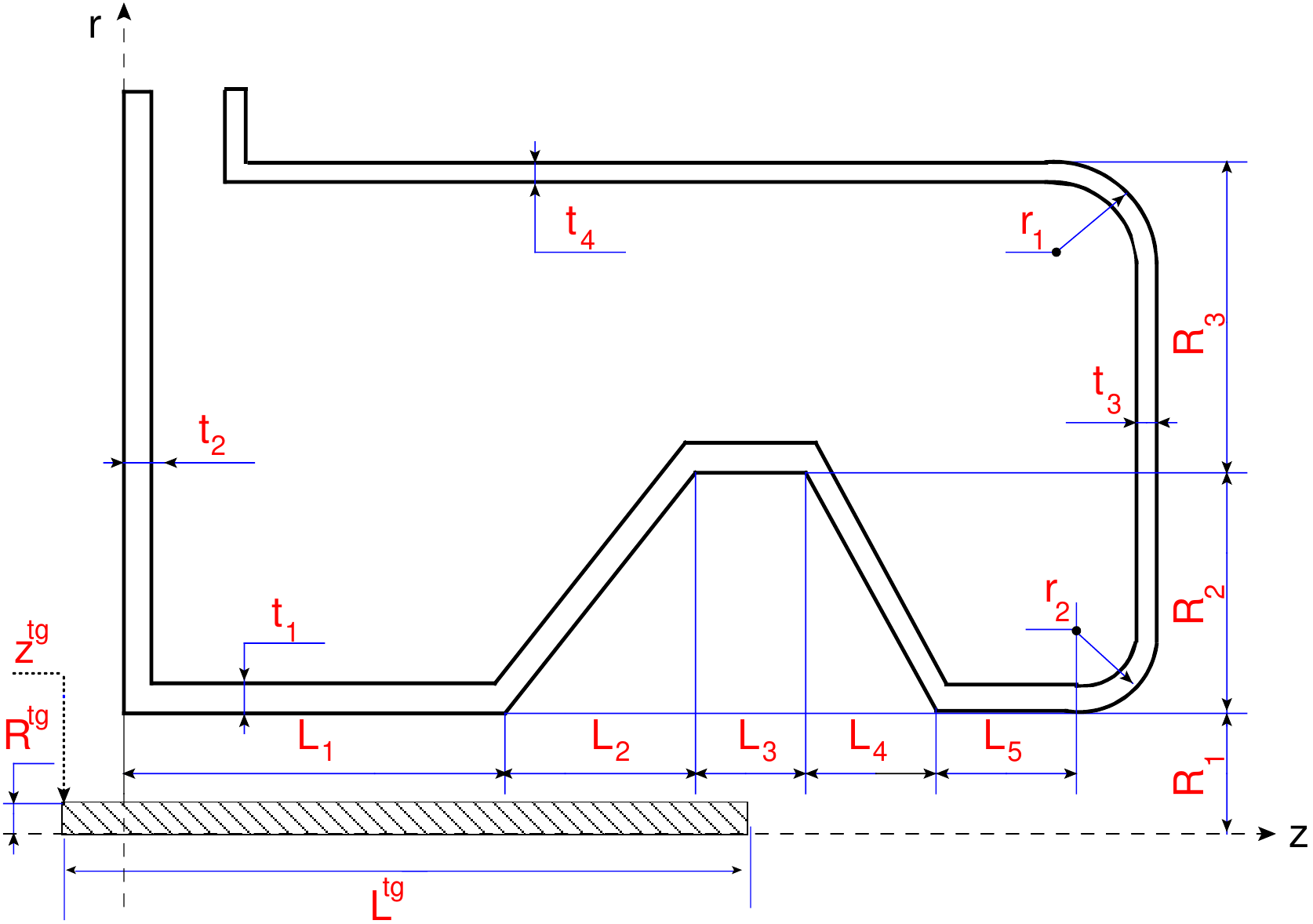}
}
\caption{ \label{fig:MiniBooNE:param} Parametrization of the forward--closed horn geometry. 
The system has a rotational symmetry around the $z$ axis. Figure shows a cut in the $r-z$ plane. 
}
\end{center}
\end{figure}

A new horn model
inspired by the one used for the MiniBooNE beam\cite{MINIBOONEbeam}, having a larger acceptance for forward produced pions,
proved to be quite effective with respect to the need for a reduced contamination from
wrong--charge pions.
The generic layout of the horn which we will use in the following (forward--closed horn hereafter) is shown in Fig.~\ref{fig:MiniBooNE:param}. 
The geometry has been parametrized in terms of the longitudinal displacement of the target with respect to the 
horn ($z_0^{\rm tg}$), 
the lengths of the longitudinal ($L_{1,2,3,4,5}$) and radial ($R_{1,2,3}$) sections, two curvature radii ($r_{1,2}$)
and the conductor thicknesses ($t_{1,2,3,4}$).

We consider two scenarios for the horn/target system: 
\begin{itemize}
\item Integrated conducting target. The target could be an integral part
of the inner horn conductor allowing for higher magnetic fields at the target exit ($R_1+t_1=R^{tg}=1.5$ cm). 
Such a setup was studied in \cite{Bobeth:2010:WP2:intTarget} taking into account 
energy deposition in the target due to the primary beam and the resistive (Joule) losses caused by the
horn current which passes the common target--inner conductor region. 
\item A target separated from the horn.
We will present how the physics performances vary when the horn inner radius is increased.
In this case the cooling system could be realized
with a helium based annular duct flow cooling as done in T2K \cite{Nakadaira:2008zz}.
\end{itemize}

Taking advantage of the small transversal dimensions of the horn,
the idea of using a battery of four horns in parallel has been
proposed. This arrangement would imply reduced stress on the targets via
lower frequency pulsing (12.5 Hz). This choice would bring 
the beam power on each target to 1 MW
which is currently considered as a viable upper limit for
solid targets operations. 

\subsection{Neutrino flux simulation}\label{sect:fluxescalc}
The neutrino energy spectra are calculated using a probabilistic approach in
order to obtain reliable results in a reasonable amount of time using samples 
of $\sim$ $10^6$ simulated protons.
The probability that the neutrino will reach the far detector is calculated at each
particle decay yielding neutrinos with analytic formulas
\cite{Campagne:2004wt,Campagne:2006yx,Blondel:2000ph,Cazes:1900zz}. The probability is then 
used as a weight factor in the calculation of the neutrino energy spectrum.
The neutrino sources considered in the simulation 
are:
\begin{enumerate}
\item Two body decays of charged pions and kaons:\\ \mbox{\{$\pi^+, K^+\} \to \ell^+ \nu_{\ell}$ and charged conjugate (c.c.)} with $\ell = \mu, e$.
Given a parent traveling in the laboratory frame with an angle $\alpha$
with respect to the beam axis, the probability for the decay $\nu$ to 
be emitted through a surface of area $A$ at a distance $L$
is:
\begin{eqnarray*}
{\mathcal{P}}_2=\frac{A}{4\pi L^2}\frac{1-\beta^2}{(\beta\cos\alpha-1)^2}
\label{eq:2body}
\end{eqnarray*}
where $\beta$ and $\gamma$ represent the relativistic factors of the parent $\pi$ or $K$ in the
laboratory frame (LF).
In this approximation, the decay region is seen as point-like from the far detector at the distance $L$.
\item Muon decays: $\mu^+ \to e^+ \nu_e \bar{\nu}_\mu$ + c.c..
The differential probability for the $\bar{\nu}_\mu$ and $\nu_e$ to be emitted parallel to the beam axis 
with an energy $E_\nu$ in the LF is:
\begin{eqnarray*}
\frac{d{\mathcal{P}}}{dE_\nu}=
\frac{2(f_0-\Pi_\mu^Lf_1\cos\theta^*)}{m_\mu\gamma_\mu(1+\beta_\mu\cos\theta^*)}
\cdot \mathcal{P}_2(\mu)
\end{eqnarray*}
where 
$\theta^*$ is the angle between the $\nu$ and the $\mu$ directions in the $\mu$ rest frame (RF). 
Denoting the $\nu$ energy in the $\mu$ RF as $E_\nu^*$, $x=2E_\nu^*/m_\mu$ and 
$f_0 = 2x^2(3-2x)$, $f_1 = 2x^2(1-2x)$ for the $\bar{\nu}_\mu$,
$f_0 = f_1= 12x^2(1-x)$ for the $\nu_e$.
$\Pi^L_\mu$ is the muon longitudinal polarization:
$ \Pi^L_\mu = \sqrt{1-(\frac{\gamma_{p}\beta_{p}}{\gamma_\mu\beta_\mu}\sin\theta^*_\mu)^2}$
where 
$\theta^*_\mu$ is the angle of the $\mu$ with respect to the beam axis in the parent ($p=\pi$ or $K$) RF.
The sign in front of $\Pi^L_\mu$  becomes $+$ for $\mu^-$ decays.
This differential probability is integrated in $E_\nu$ sampling the energy which is available for the $\nu$ 
uniformly in bins of 1 MeV width.
In order to reduce the statistical fluctuation which would arise from the limited number of muons 
decaying in the decay tunnel ($\beta\gamma c \tau \simeq 6.3$~km for $p_\mu=1$~GeV/c) each time a $\mu$ 
is produced the probability of its decay is introduced as a weight. This is calculated from the length traveled
inside the decay tunnel assuming a straight line propagation.

\item Kaons semi-leptonic three body decays: $K \to \pi \ell \nu_\ell$ with $\ell = \{\mu, e\}$ and
  $K = \{K^\pm, K^0\}$. The probability to reach the detector is computed as: 
\begin{eqnarray*}
\frac{d{\mathcal{P}}}{dE_\nu}=
\frac{2 f(E_\nu^*)}{(m_K-m_\pi-m_l)\gamma_K(1+\beta_K\cos\theta^*)}\cdot \mathcal{P}_2(K)
\end{eqnarray*}
$\theta^*$ is the angle between the $\nu$ and the $K$ directions in the $K$ RF.
This differential probability is integrated in $E_\nu$ sampling the energy which is available for the $\nu$ 
uniformly with a fixed number of 200 points. $f(E_\nu^*)$ is a 
parametrization of the distribution of the energy of the $\nu$ in the $K$ RF
which was derived from \cite{Kformfact}.
\end{enumerate}

The distribution of the secondaries at target exit obtained with the FLUKA \cite{Battistoni:2007zzb} generator
is used as an external input to a GEANT4 \cite{GEANT4} simulation derived from a  GEANT3 code developed in \cite{Campagne:2004wt}.
The target, the horn with its magnetic field and the decay tunnel are fully simulated 
within GEANT4. Alternatively GEANT4 can be used to simulate also the interactions of primary protons in the target:
this option was used as cross check (Sect. \ref{sect:syte}).
In order to cross-check and validate the new GEANT4--based software, a
comparison has been done with the fluxes obtained with GEANT3. 
The fluxes obtained in the two frameworks are in good agreement both in terms of normalization and shape \cite{NUFACT09}.
Further cross-checks included the correct implementation of the decay branching ratios,
a comparison with an independent code and a check based on direct scoring of the emitted neutrinos.

\subsection{Sensitivities}
\label{sect:limits}
The sensitivities for the measurement of the oscillation parameters
$\theta_{13}$ and $\delta_{CP}$ are obtained with the help of GLoBES 3.0.14
\cite{Huber:2004ka}.
The parametrization of the MEMPHYS Water Cherenkov detector\cite{deBellefon:2006vq} at the level of
physics performance (efficiencies, background rejection, etc.) is
implemented in the public AEDL file \verb+SPL.glb+ which is distributed with
GLoBES. The results derive from the analysis algorithms developed in the context of the 
SuperKamiokande detector. The efficiency for the reconstruction of $\nu_e$ events is 70\%, the
fraction of misidentified muon events is 0.054\% of the $\nu_\mu^{CC}$ events and the
neutral current $\pi^0$ background is 0.065\% of the un-oscillated $\nu_\mu^{NC}$ and 0.25\% of the un-oscillated 
$\bar{\nu}_\mu^{NC}$.
A detailed description can be found in \cite{Campagne:2006yx}. 
In the following we assume a sharing between neutrino and anti-neutrino 
running of 2 and 8 years, respectively, as it was done in \cite{Campagne:2006yx}.
A systematic error of 5\% on the fluxes is assumed in the calculation of the limits.

The sensitivity limit is defined by generating the event rates in the assumption of a null $\theta_{13}$
and performing a fit with finite values of $\theta_{13}$. In particular, in the optimization process (Sect. \ref{sect:OPT}), 
we assumed the normal hierarchy and the $\Delta\chi^2$ was calculated neglecting the correlations 
of the parameters with the \verb+glbChiSys+ routine (see \cite{Huber:2004ka}). 

\section{Sensitivity-based optimization
\label{sect:OPT}}
The approach which was followed in the optimization of the forward--closed horn and 
the decay tunnel uses the final $\sin^2 2\theta_{13}$ sensitivity, i.e. the final physics performance as a
guiding principle in the ranking of the configurations under scrutiny\footnote{The possibility to use the sensitivity 
to CP violation as a guiding line will be the object of a future study.}. In the evaluation
of this quantity a complex set of relevant factors are given as an input: the normalization and shape 
of each neutrino flavor, the running time in the positive and negative focusing mode, 
the energy dependence of the cross sections, the backgrounds in the far detector and its response
in terms of efficiency and resolution.

Given the well known dependence on the $\sin^22\theta_{13}$ limit on the $\delta_{CP}$ phase, we introduced
the quantity $\lambda$ defined as the $\delta_{CP}$-averaged 99\% C.L. sensitivity limit on
$\sin^2 2\theta_{13}$ 
in units of $10^{-3}$:
\begin{equation}
  \label{eq:lambdadef}
  \lambda = \frac{10^3}{2\pi} \int_{0}^{2\pi} 
     \lambda_{99}(\delta_{CP})\, d\delta_{CP}
\end{equation}
This quantity has been used as a practical way of defining with a single number the 
quality of the focusing system. 

The key parameters defining the horn and tunnel geometry are randomly sampled within specified 
ranges and the correlations with the figure of merit $\lambda$ studied. 

\subsection{Beam composition and achievable limits}

We started by studying the correlation between the relative normalizations of $\nu_\mu$ and $\nu_e$ fluxes and
the corresponding variations of $\lambda$. For this exercise
we chose a specific horn configuration providing a $\lambda \simeq 1$ 
and scaled the $\nu_\mu$, $\bar{\nu}_\mu$ fluxes 
globally by a factor $c_\mu$ with $0.5<c_\mu<2$. The same scaling was applied to the $\nu_e$ and $\bar{\nu}_e$
fluxes 
with an independent factor $c_e$ within the same limits.

The parameter $\lambda$ improves (i.e. decreases) by a factor of about 3 when the
$\nu_e$ is reduced by a factor 2 and $\nu_\mu$ increased by the same factor
while it worsens by a factor 2.4 when the variation of the fluxes
is done in the other sense.
The iso-sensitivity levels follow quite closely a law $c_\mu = \sqrt{c_e}$.
This behavior reflects the fact that the experiment significance 
$\rm{S}/\sqrt{\rm{B}}$ in terms of observed events in the far detector, to which $\lambda$ is related,
is approximately\footnote{
The signal (S), the $\nu_e$ appearance, is a fixed fraction of the
$\nu_\mu$ flux and the background (B) is mostly generated by the
intrinsic $\nu_e$ component in the beam.} invariant for a scaling of the fluxes ($c_\mu$, $c_e$) 
in which $c_\mu = \sqrt{c_e}$.  

We have studied the correlation between the integral fluxes $\nu_\mu$ and
$\nu_e$ with the full simulation under variations of the decay tunnel geometry only
with the horn configuration which will be described in Sect. \ref{pippo}. The correlation is strong
since the bulk of the $\nu_e$ are generated by the same decay chain producing $\nu_\mu$
($\pi^+\to\mu^+\nu_\mu$, $\mu^+\to e^+\nu_e\bar{\nu}_\mu$ + c.c.).
The dependence follows roughly a quadratic law: $\phi(\nu_e) \sim \phi(\nu_\mu)^\alpha$, $\alpha\simeq 2$.
Deviations occur especially at small fluxes ($\phi_+(\nu_\mu) \simeq 3.5 \cdot 10^{14}$ $\nu/100\rm{m}^2/\rm{year}$ at 100~km) 
below which $\alpha<2$. This sets approximately the threshold above which increasing
the $\nu_\mu$ flux by tuning the decay tunnel does not improve the sensitivity.

\subsection{The optimization procedure}
\label{pippo}
The parameters of the forward--closed horn and of the decay tunnel were sampled with uniform 
probability distributions imposing the configuration to be  geometrically consistent 
(``iteration-1''). The decay tunnel length ($L^{\rm{tun}}$) and radius ($R^{\rm{tun}}$) were restricted 
in the intervals $[35, 45]$~m and $[1.8,2.2]$~m respectively.
The maximal length and radius of the horn were limited
to 2.5~m and 80~cm in order to maintain a compact design which
allows to use a battery of four target-horn stations in parallel.
Moreover, the inner radius $R_1$ was limited in [1.2, 4]~cm, the lower limit corresponding to the ``integrated target'' limit. 
Further constraints on the parameters were $L_1>50$~cm, $L_5<15$~cm and $-30<z_0^{\rm{tg}}<0$~cm.

The stability of the value of $\lambda$ related to statistical
fluctuations ($2\cdot10^5$ tracks per configuration) was estimated by repeating the simulation using an identical set of
parameters with independent track samples. 
We find a r.m.s. of the integral $\nu_\mu$ and $\nu_e$ fluxes respectively at the level of 3\% and 5\% of the central value. 

\begin{figure}
\centering
\resizebox{0.5\textwidth}{!}{
\includegraphics{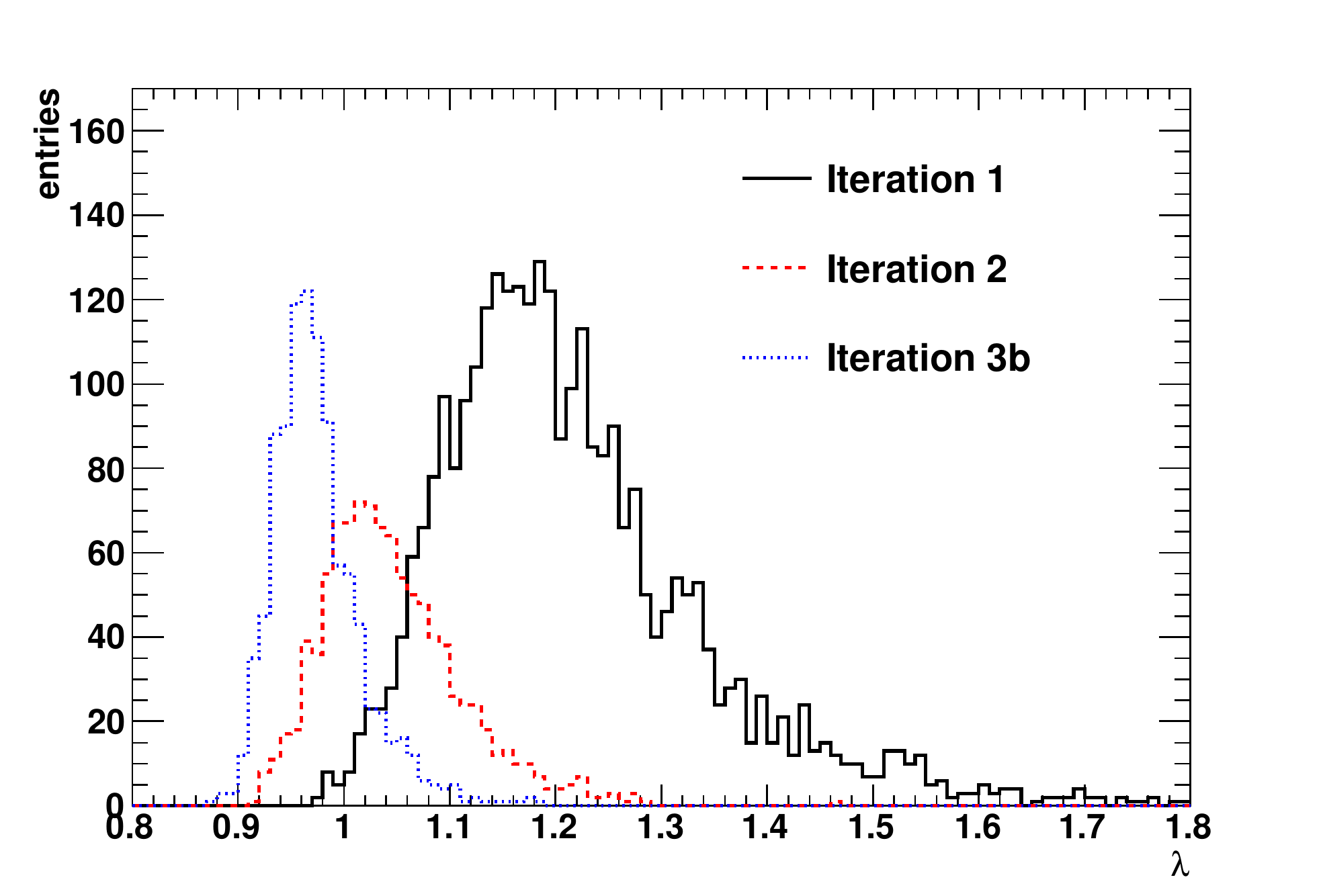}
}
\caption{ \label{fig:conf_sup} Distribution of $\lambda$ for iteration-1 (continuous, 3000 configurations), iteration-2
  (dashed 1000 configurations), and iteration-3b (dotted, 1000 configurations). See the text for the definition of the samples.}
\end{figure}

The distribution
of $\lambda$ is shown in the continuous histogram of Fig.~\ref{fig:conf_sup}. 
Studying the distributions of the integral fluxes for positive and negative focusing 
for the sub-sample ($\mathcal S$) of configurations providing good limits ($\lambda<1.05$) we 
observe an enrichment in configurations yielding high $\nu_\mu$ fluxes with a low $\nu_e$ and 
$\bar{\nu}_\mu$ contamination as expected. Small  $\nu_e$ contamination in the $\bar{\nu}_\mu$
flux is confirmed to be also important (see Sect. \ref{sect:horn}).
On the other hand the mean and the root mean square of the $\nu_\mu$ energy spectrum do not
show a visible correlation with $\lambda$.

\begin{figure}
\begin{center}
\resizebox{0.5\textwidth}{!}{
\includegraphics{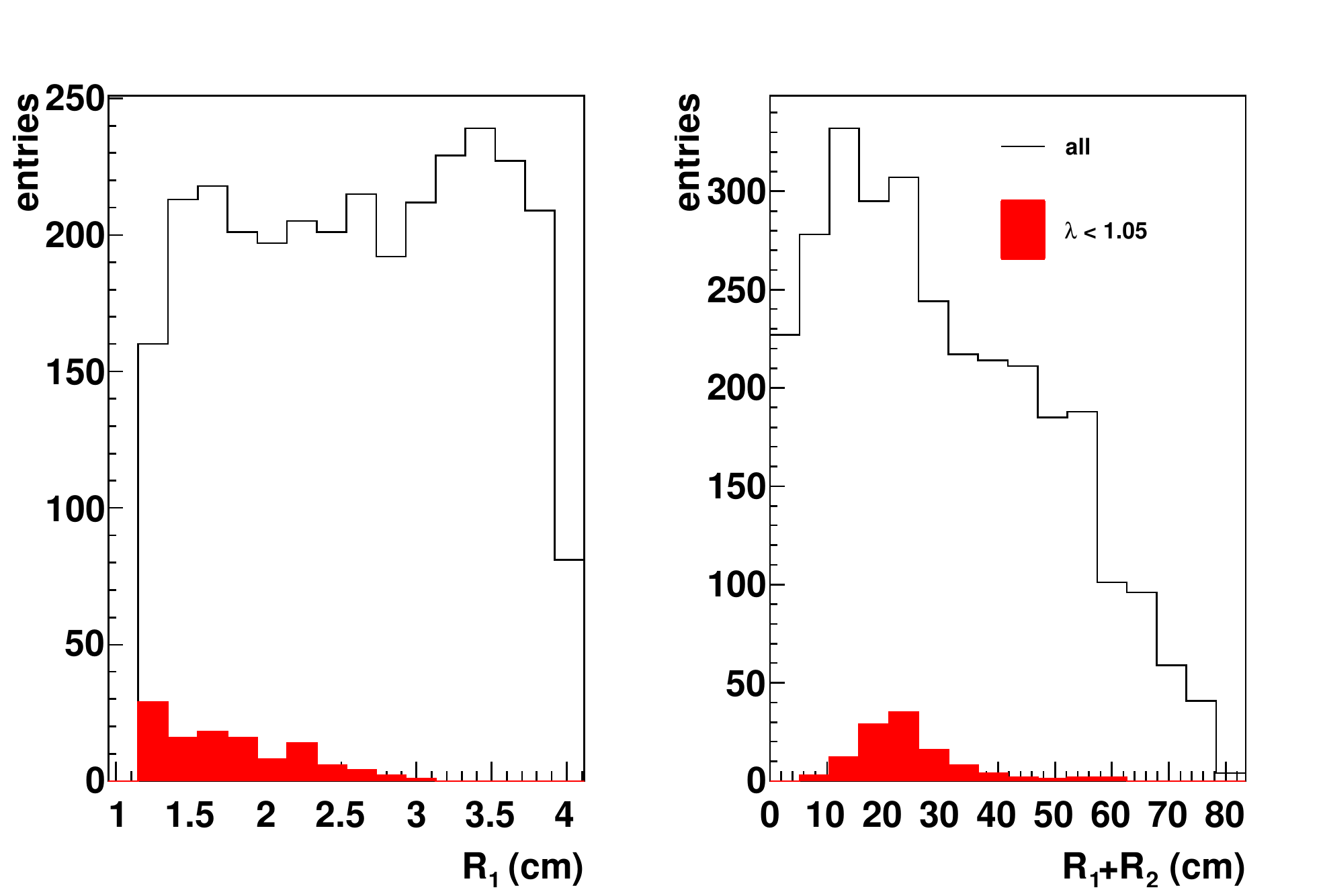}
}
\caption{ \label{fig:plotconfs5} 
Distribution of the horn geometrical parameters $R_1$ (inner radius) and $R_1+R_2$ (alcove radius) 
for the inclusive sample (empty histogram) and for the $\mathcal{S}$ subsample (filled histogram). See text.}
\end{center}
\end{figure}

The statistical distributions of the geometrical parameters 
for the inclusive sample and the $\mathcal{S}$ subsample were compared
to obtain an indication of the parameters which are more effective in increasing the sensitivity. Despite
the averaging on a large set of parameters variations, the distributions show interesting features
for the following parameters: the horn inner radius ($R_1$), which has
a strong preference for small values; the target position ($z_0^{\rm tg}$), for which very low values are disfavored; 
the radius of the ``alcove'' ($R_1+R_2$) whose optimal values cluster at about 20~cm and
the decay tunnel length ($L^{\rm tun}$) which tends to give better performance for small values. 
Two of these parameters ($R_1$ and $R_1+R_2$) are shown as an 
example in Fig.~\ref{fig:plotconfs5}. 

\begin{figure}
\centering
\resizebox{0.4\textwidth}{!}{
\includegraphics{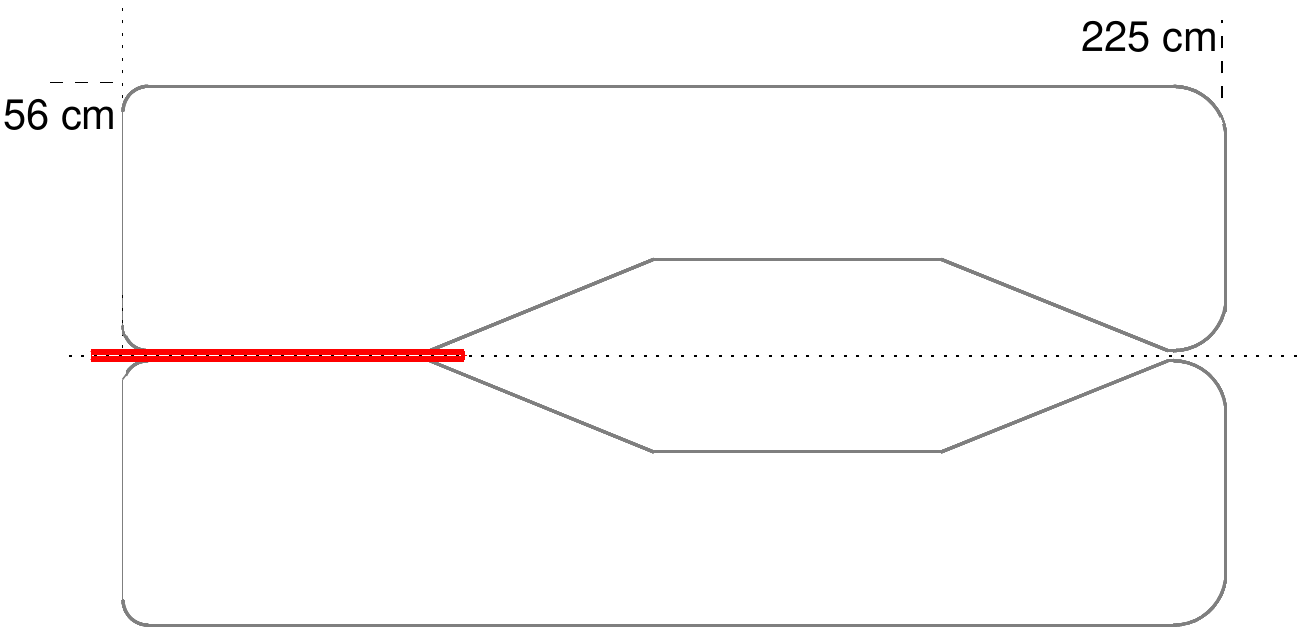}
}
\caption{ \label{fig:theone} Shape of the optimal horn.}
\end{figure}

According to these indications a second scan (``iteration-2'') was performed
after fixing the parameter $R_1$  at 1.2~cm and restricting the
ranges of variation:
   $ 20<R_1+R_2<22 ~\rm{cm}$,
   $30<L^{\rm tun}<40~\rm{m}$, 
   $-15<z_0^{\rm tg}<0~\rm{cm}$.
The distribution of $\lambda$ for this new sample is shown by the dashed histogram of Fig.~\ref{fig:conf_sup}. 
With respect to the same distribution for the previous scan (continuous line) 
a shift towards better limits by about 20\% is achieved. 
The horn shape for the configuration giving the minimum value for $\lambda$ 
in the iteration-2 is shown in Fig.~\ref{fig:theone}. 

\begin{table}
\begin{center}
\begin{tabular}{|c|c||c|c|}
\hline
$L_1$ & 58.9&$r_1=r_2$ & 10.8\\
$L_2$ & 46.8&$R_1$ & 1.2\\
$L_3$ & 60.3&$R_1+R_2+R_3$ & 56.2\\
$L_4$ & 47.5&$R_1+R_2$ & 20.3\\
$L_5$ & 1.08&$z_{0}^{\rm tg}$ &-6.8\\
\hline
$L^{\rm tg}$ & 78 &$R^{\rm tg}$ &1.5\\
\hline
$L^{\rm tun}$ & 2500 &$R^{\rm tun}$ &200\\
\hline
\end{tabular}
\end{center}
\caption{Parameters of the optimized system expressed in cm.}
\label{tab:besth}
\end{table}
Finally we kept the horn shape fixed
(Fig.~\ref{fig:theone}) and performed a further tuning of the decay tunnel length and radius. 
The scan was done in two regions to better constrain the position of the minimum:
$0.5<R^{\rm tun}<2.5$ m and $10<L^{\rm tun}<60$ m
(``iteration-3a'') and 
$1.5<R^{\rm tun}<4.5$~ m and $15<L^{\rm tun}<35$~m (``iteration-3b''). 
The distributions of the parameter $\lambda$ for the latter sample, 
which lies in the neighborhood of the minimum, is shown by the dotted 
histogram of Fig.~\ref{fig:conf_sup}. With respect to the initial distribution an improvement 
of 25-30\% is obtained.

\begin{figure}[hpt!]
\centering
\resizebox{0.49\textwidth}{!}{
\includegraphics{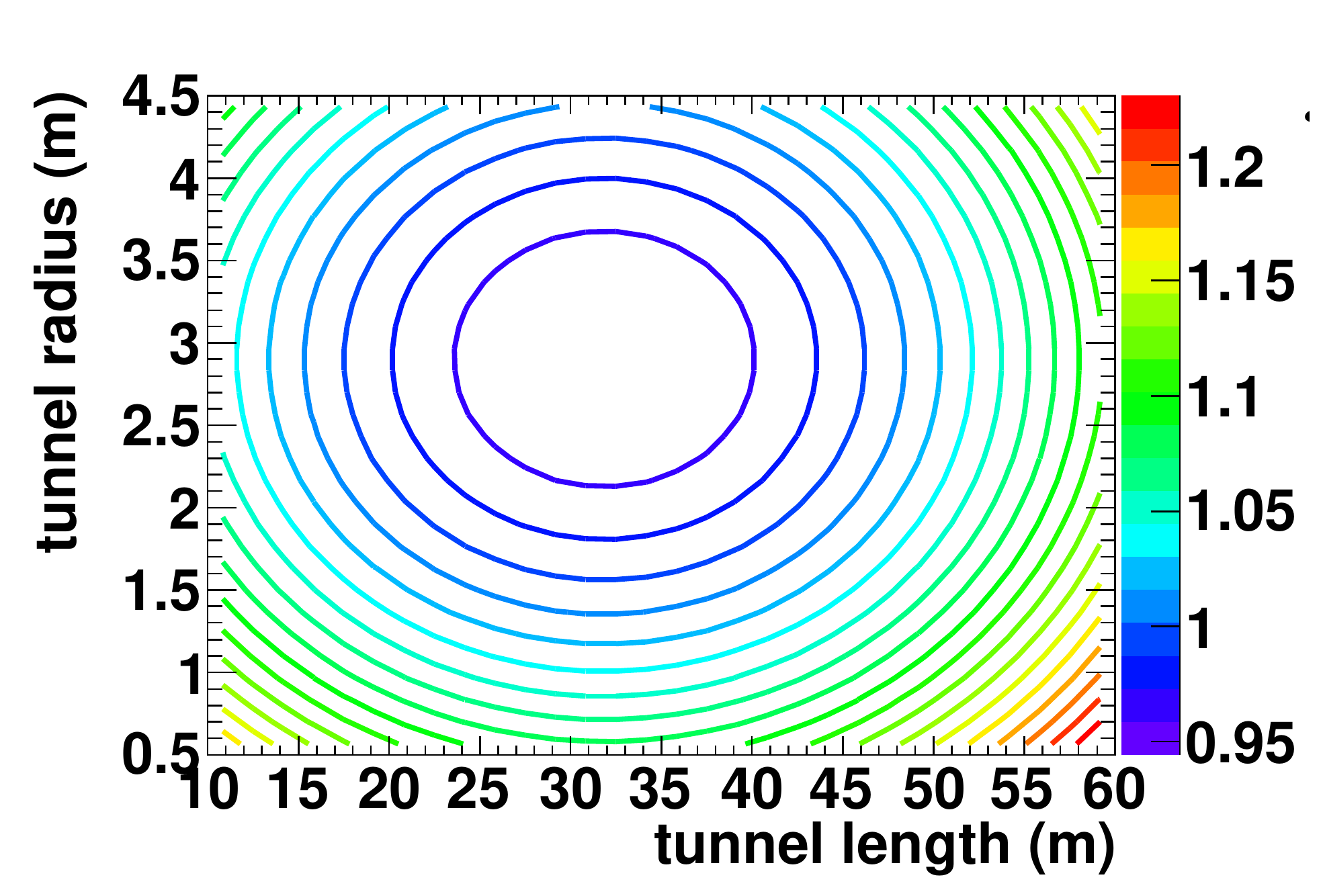}%
\includegraphics{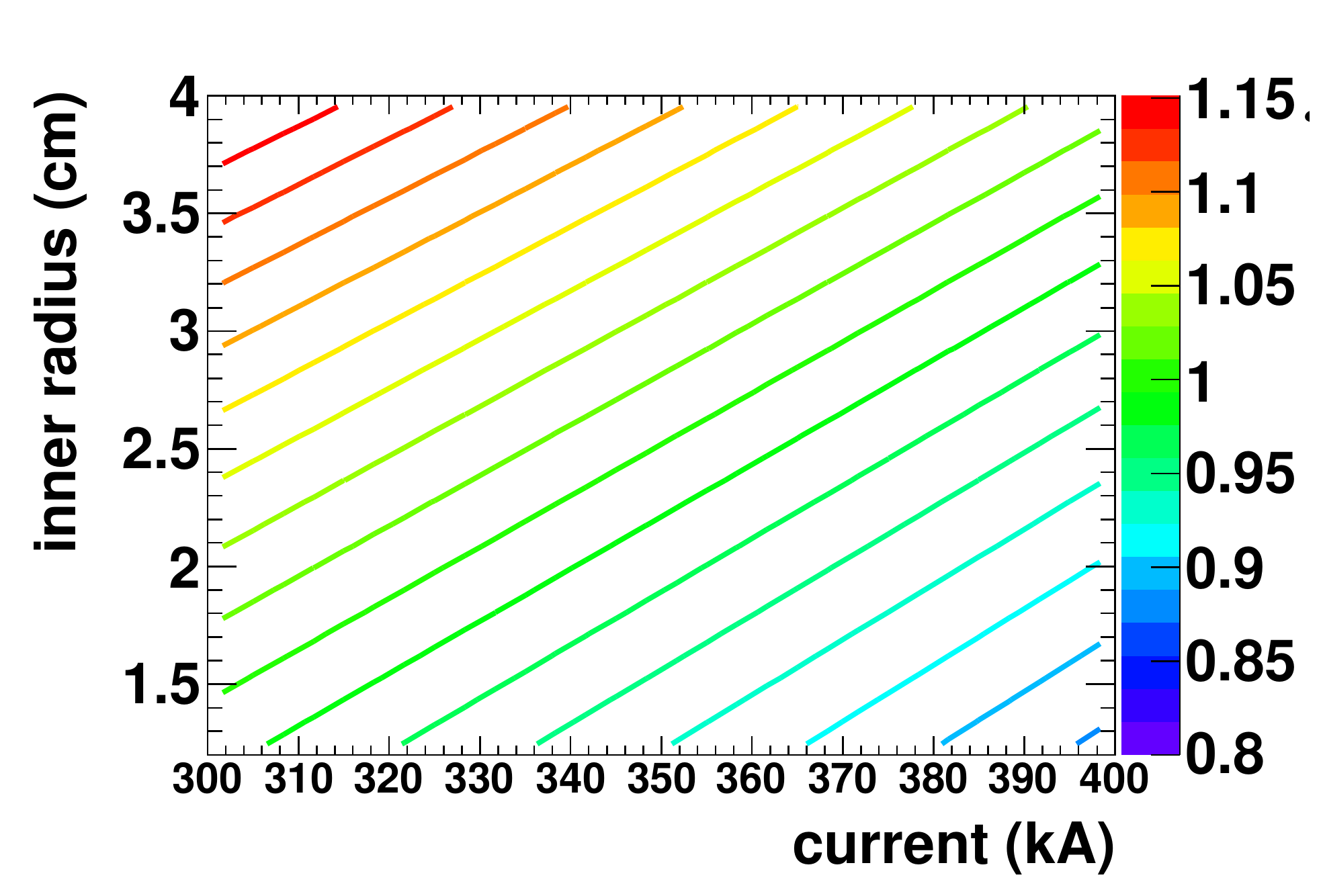}
}
\caption{ \label{fig:Ltun} Dependence of $\lambda$ in the $L^{\rm tun}$ - $R^{\rm  tun}$ plane (left)
and in the ($R_1$, $i$) plane (right). 
 }
\end{figure}

The dependence of $\lambda$ on the decay tunnel variables can be reasonably fitted
with a quadratic function:
$\lambda=0.94+2.1\cdot 10^{-4}(L^{\rm{tun}}[{\rm{m}}]-31.8)^2+2.4\cdot 10^{-2}(R^{\rm{tun}}[\rm{m}]-2.9)^2$
(Fig.~\ref{fig:Ltun}, left).
Since the minimum is relatively broad we chose  
$L^{\rm tun} = 25$~m  and $R^{\rm tun} = 2$~m as central values based on practical considerations related 
to the excavation and shielding of large volumes. This compares to the previous values of $40$ m
of length and 2 m of radius.

We have also observed that an increase in the current (between 300 and 400~kA)
tends to systematically produce better sensitivity limits. 
The interplay between the current and the inner horn radius in terms of the quantity $\lambda$, 
after having fixed the optimal horn and decay tunnel has been studied.
Data are well fitted by a linear function in two dimensions:
$\lambda=(9.2-0.81\cdot I [100~\rm{kA}])/(7.3-0.37\cdot R_1[\rm{cm}])$ 
(right-hand plot of  Fig.~\ref{fig:Ltun}).
The effect of increasing the current, i.e. a stronger magnetic field in the vicinity of the target, is physically equivalent to
decreasing the minimum horn inner radius. In this way, at first approximation, working at constant $I/R_1$ ($\propto B$), 
allows to stay at fixed sensitivity. 

In the following we will take the configuration with a current of 300~kA and integrated target as the baseline choice.
It can be noticed however that, in the case of a non-integrated target, if the minimum horn radius would need 
to be extended up to 4~cm to accomodate for the cooling system, similar performances could be recovered 
by increasing the current up to about 400~kA (right-hand plot of  Fig.~\ref{fig:Ltun}).

\section{Results with the optimized setup}
\label{sect:resOPT}

\begin{figure}
\begin{center}
\resizebox{0.5\textwidth}{!}{
\includegraphics{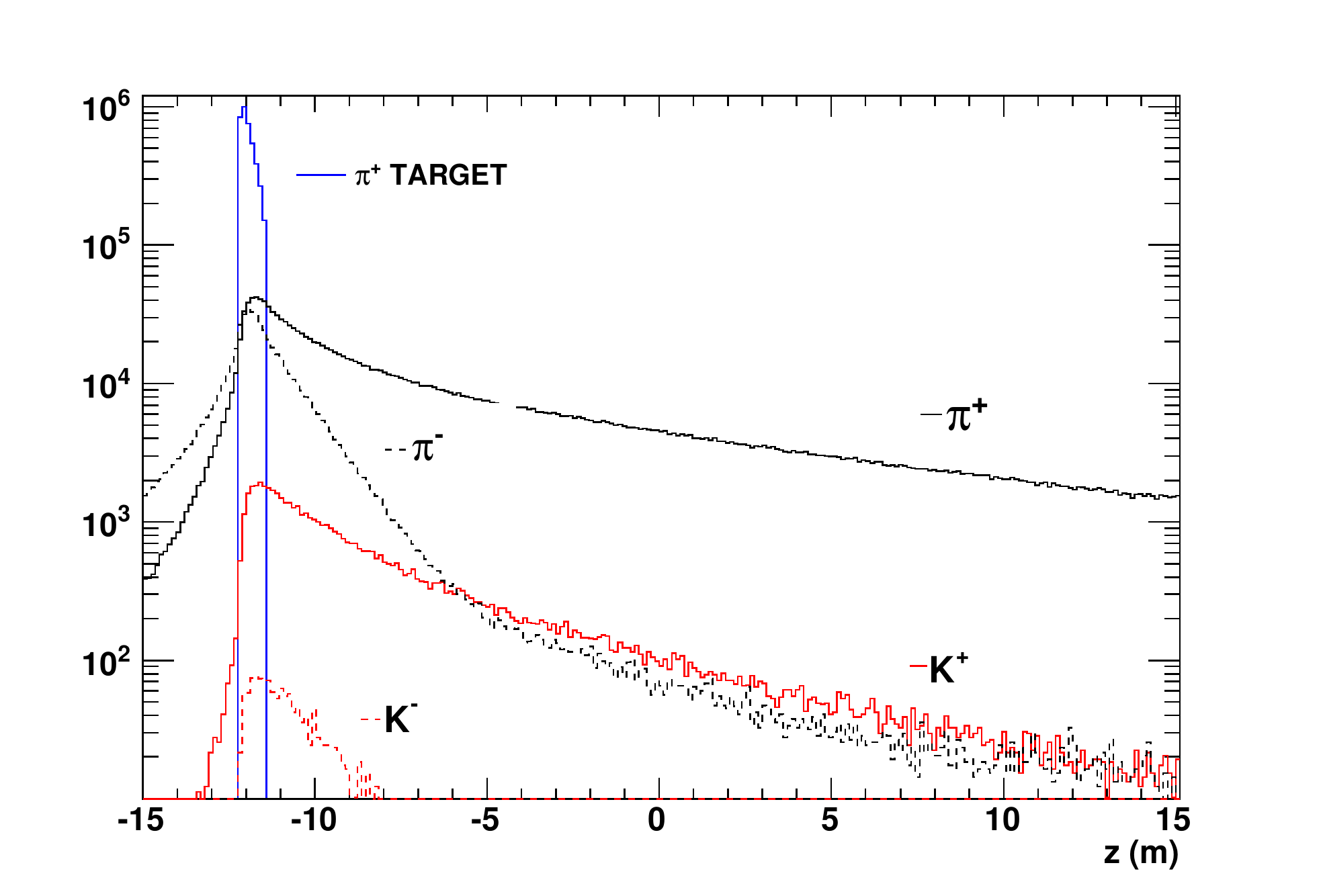}
}
  \caption{Distributions of the longitudinal coordinate ($z$) of the decays in flight of positive and negative pions and kaons 
with the optimized configuration.}
 \label{fig:DIF}
\end{center}
\end{figure}

The distribution of the longitudinal coordinate ($z$) of the decays in flight of charged pions and kaons 
with the optimized configuration in the positive focusing mode is shown in Fig.~\ref{fig:DIF}. The continuous histogram on the left-hand side represents the $z$ distribution of the exit point 
of positive pions from the target. The focusing effect is visible in the different shape observed for positive and negative mesons. 
 The relative yield and lifetime of pions and kaons can be also appreciated. 

\begin{figure}
\begin{center}
\resizebox{0.5\textwidth}{!}{
 \includegraphics{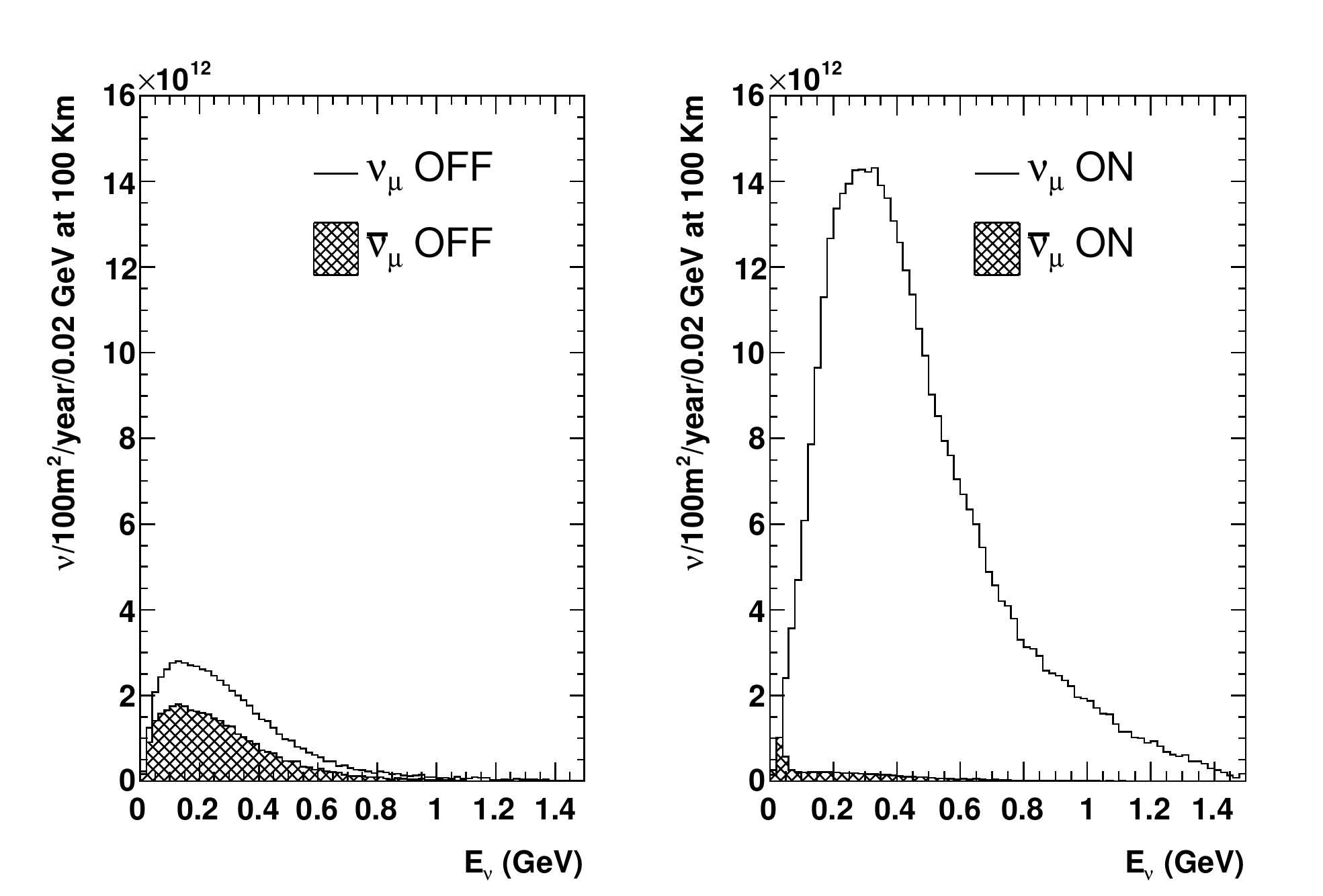}
}
  \caption{Effect of the focusing. $\nu_\mu$ and $\bar{\nu}_\mu$ spectra
  in the positive focusing mode with horn current off (left) and on (right).}
 \label{fig:ONOFF}
\end{center}
\end{figure}

The focusing effectiveness of the system can be characterized by comparing the $\nu_\mu$
and $\bar{\nu}_\mu$ spectra at the far detector obtained with null current and the nominal one.
Working in the positive focusing mode the $\nu_\mu$ flux
is increased of a factor 6.5 and the $\bar{\nu}_\mu$ flux decreased by a factor 5.4 with respect to 
the situation in which no focusing would be applied (Fig.~\ref{fig:ONOFF}).

\begin{figure}
\centering
\resizebox{0.5\textwidth}{!}{
\includegraphics{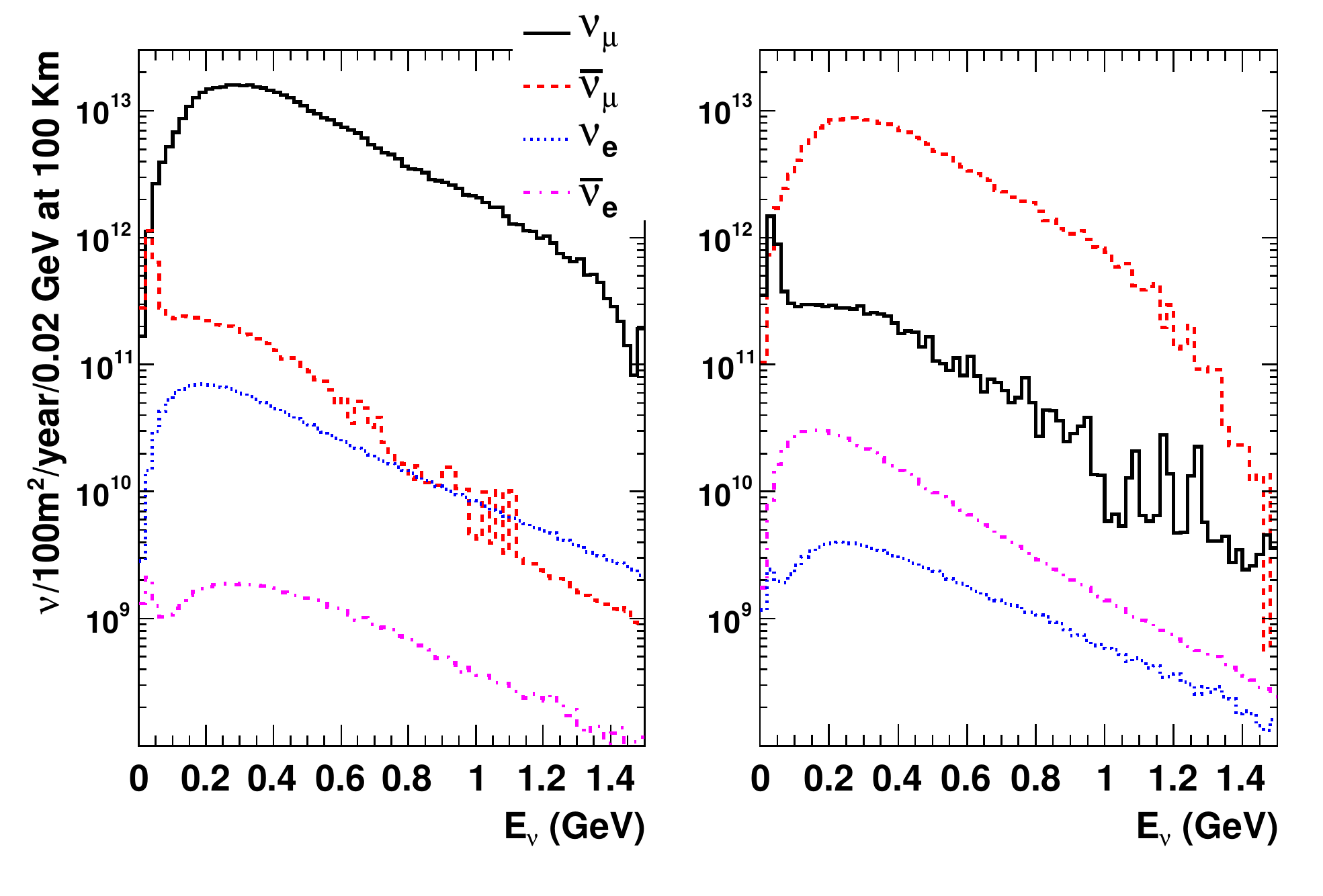}
}
\begin{tabular}{|c|c|c|c|c|}
\hline
focusing &$\nu_\mu$ & $\bar{\nu}_\mu$ & $\nu_e$ & $\bar{\nu}_e$\\ 
\hline
+ & $3.9\cdot 10^{14}$ & $6.3\cdot 10^{12}$ & $1.7\cdot 10 ^{12}$ & $6.0\cdot 10^{10}$ \\
&\scriptsize{98.0\%}& \scriptsize{1.6\%}& \scriptsize{0.42\%}& \scriptsize{0.015\%}\\
\hline
- & $1.0\cdot 10^{13}$ & $2.2\cdot 10^{14}$ & $1.2\cdot 10 ^{11}$ & $6.4\cdot 10^{11}$ \\
&\scriptsize{4.4\%} & \scriptsize{95.3\%} &\scriptsize{0.05\%} &\scriptsize{0.28\%}\\
\hline
\end{tabular}
\caption{Neutrino fluxes obtained with the optimized horn and decay tunnel
 in positive (left) and negative (right) focusing mode.  The table contains the integral neutrino flux per year for each flavor at 
a distance of 100~km over a surface of 100~m$^2$. The fluxes were obtained with a sample of $10^7$ simulated proton-target interactions.}
\label{fig:fluxesFC}
\end{figure}

The obtained $\nu_\mu$, $\nu_e$ and charged conjugate (c.c.) neutrino fluxes 
are shown in Fig.~\ref{fig:fluxesFC} for positive (left) and negative focusing (right) runs. 
They correspond to 5.6~$\cdot$~10$^{22}$ protons on target (p.o.t.)/year
(4~MW $\cdot$ 10$^7$ s at 4.5~GeV) and are calculated at a reference distance of 100~km over a surface of 100~m$^2$. 
The fractions of $\nu_\mu$, $\bar{\nu}_\mu$, $\nu_e$ and $\bar{\nu}_e$ with respect to the total
are (98.0\%, 1.6\%, 0.42\%, 0.015\%) and (95.3\%, 4.4\%, 0.28\%, 0.05\%) for the positive and negative focusing 
modes respectively.
In positive (negative) focusing mode the $\nu_e$ ($\bar{\nu}_e$) 
fluxes are dominated by muon decays: 82\% (90\%). 
The c.c. fluxes receive instead a large contribution from kaon three body decays 
(81\% and 75\% in positive and negative focusing respectively)
with muon decays from the decay chain of ``wronge charge'' pions at low energy
contributing for the rest. The fluxes are publicly available \cite{fluxesWEB}.

\begin{figure}
\centering
\resizebox{0.5\textwidth}{!}{
\includegraphics{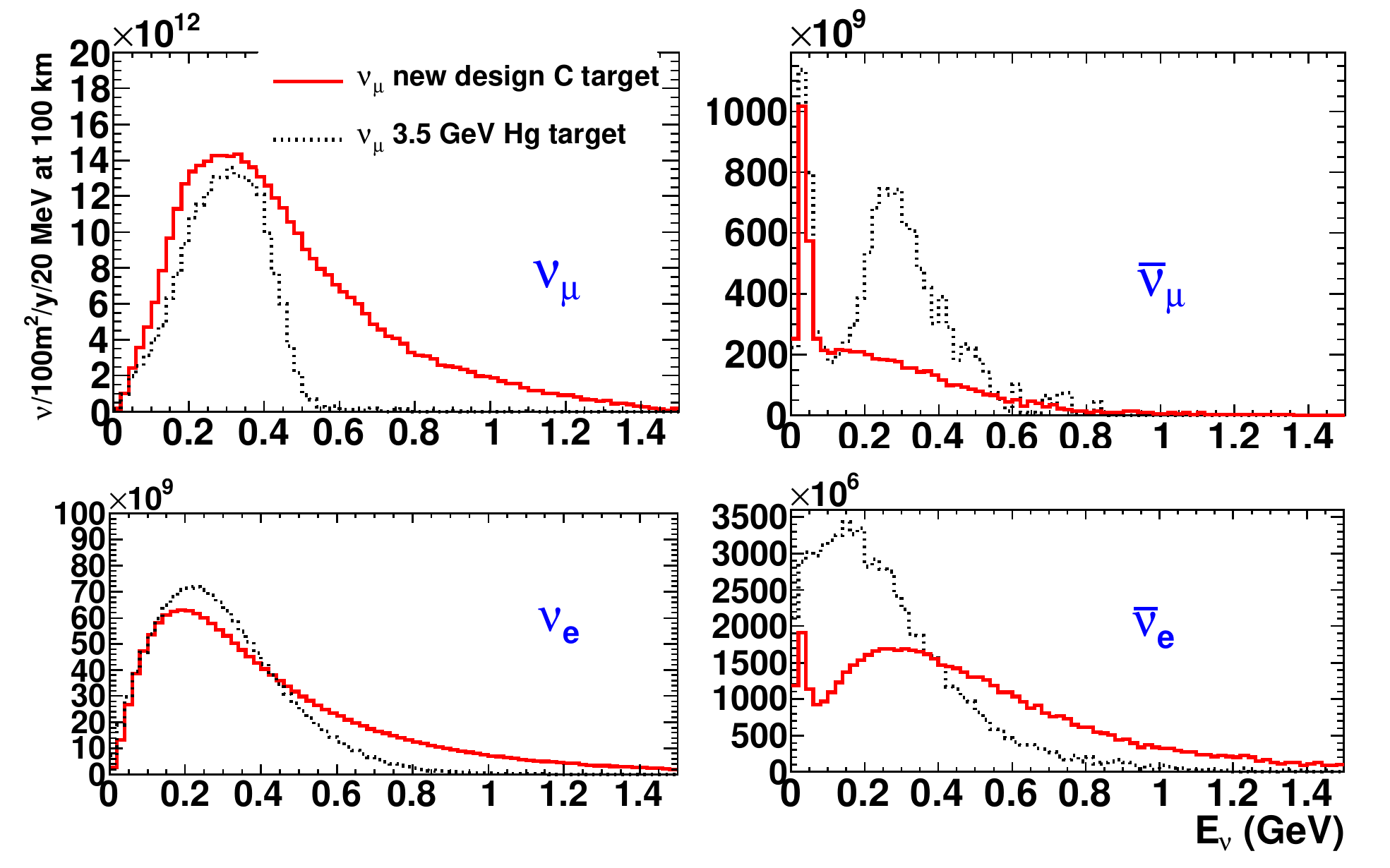}
}
\caption{Comparison of the neutrino fluxes obtained with the new design (continuous line) and the previous one \cite{Campagne:2006yx} (dotted line).}
\label{fig:fluxesNEWOLD}
\end{figure}

In Fig.~\ref{fig:fluxesNEWOLD} the fluxes obtained with the optimized horn are compared to those obtained with 
the original double conical horn with currents of 300 and 600~kA associated with a mercury target and published in \cite{Campagne:2006yx}. 
The $\nu_\mu$ and $\nu_e$ energy spectra are shifted to higher
energies with an increase in statistics particularly around 500
MeV. The $\nu_\mu$ flux is enhanced also in the proximity of the oscillation maximum at 260 MeV 
where  the $\nu_e$ flux is reduced by a similar fraction. The wrong-CP component ($\bar{\nu}_e$, $\bar{\nu_\mu}$) 
on the other hand is reduced by more than a factor two. 

\subsection{Parallel horns}
\label{sect:MHORN}
\begin{figure}
\centering
\resizebox{0.4\textwidth}{!}{
\includegraphics{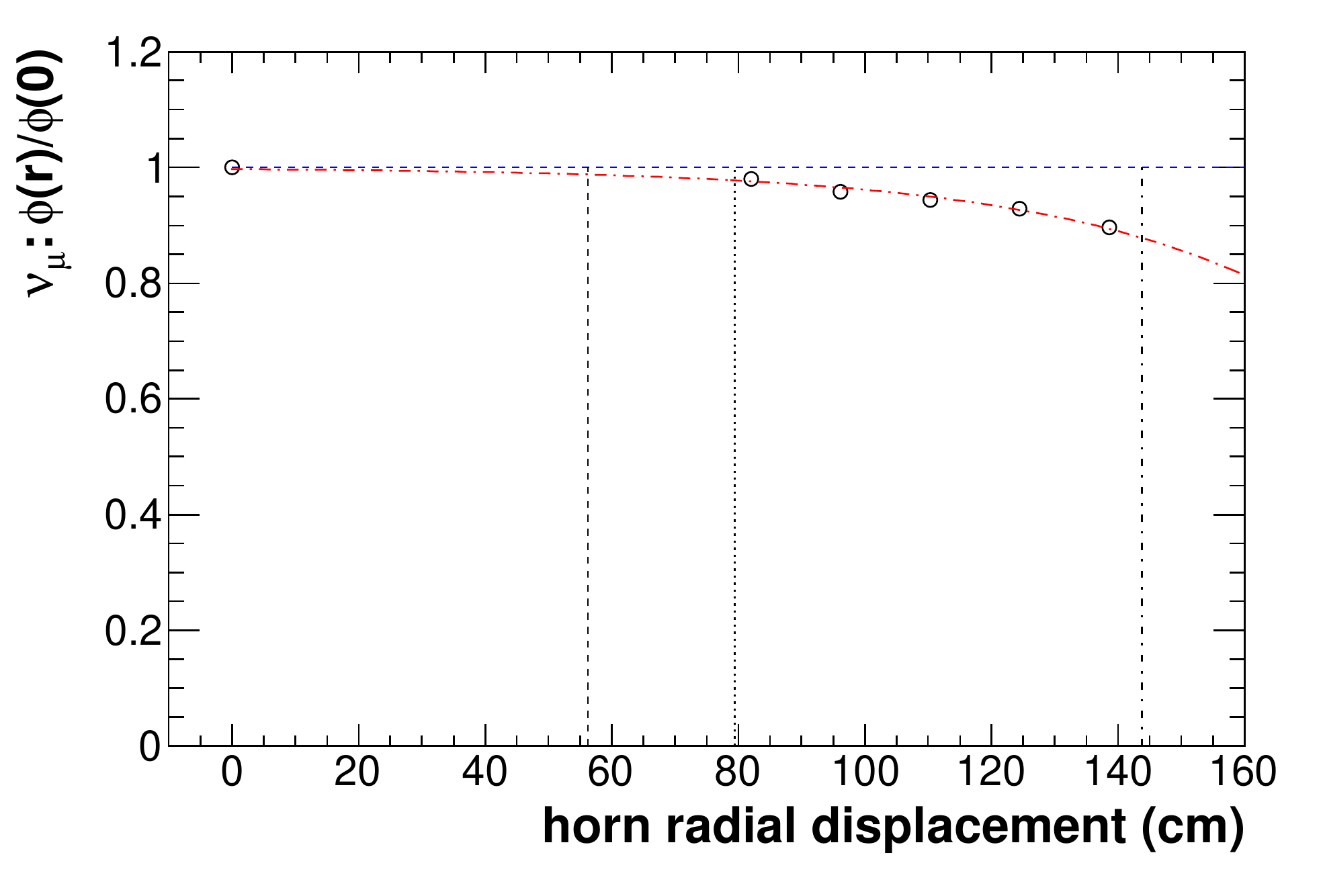}
}
\caption{Variation of the ratio of $\nu_\mu$ fluxes obtained with a  set of four displaced horns ($\phi(r)$) with respect 
to a single centered horn ($\phi(0)$) as a function of the displacement $r$. Starting from the left the first line represents
the horn radius, the second and third lines the minimum and maximum possible displacement.
}
\label{fig:line}
\end{figure}

Fig.~\ref{fig:line} shows dependence of the $\nu_\mu$ flux obtained 
using four horns as a function of their radial displacement ($r$) with respect 
to the decay tunnel axis. Even in the most pessimistic case 
where the horns should be placed at the decay tunnel edge ($r=R^{\rm{tun}}-R^{\rm{horn}}$), 
the flux reduction is only 10\%. 
The baseline configuration with horns as central as
possible (i.e. in mutual contact, $r \sim R^{\rm{horn}}\sqrt{2}$) causes a minor loss
of $\nu_\mu$ of the order of 1-2\%.  The presence of a magnetic field in all the horns
simultaneously or in each horn separately does not alter
significantly the fluxes.

\subsection{Hadroproduction uncertainties}
\label{sect:syte}
In order to characterize the subsample of produced pions giving the most
important contribution to the neutrino flux, the distribution of the momentum ($p$) and polar angle ($\theta$)
of positive pions was considered in bins of the neutrino energy. The pion distributions in each neutrino energy bin 
were averaged taking the flux in the corresponding bin as a weight. 
\begin{figure}
\begin{center}
\resizebox{0.5\textwidth}{!}{
\includegraphics{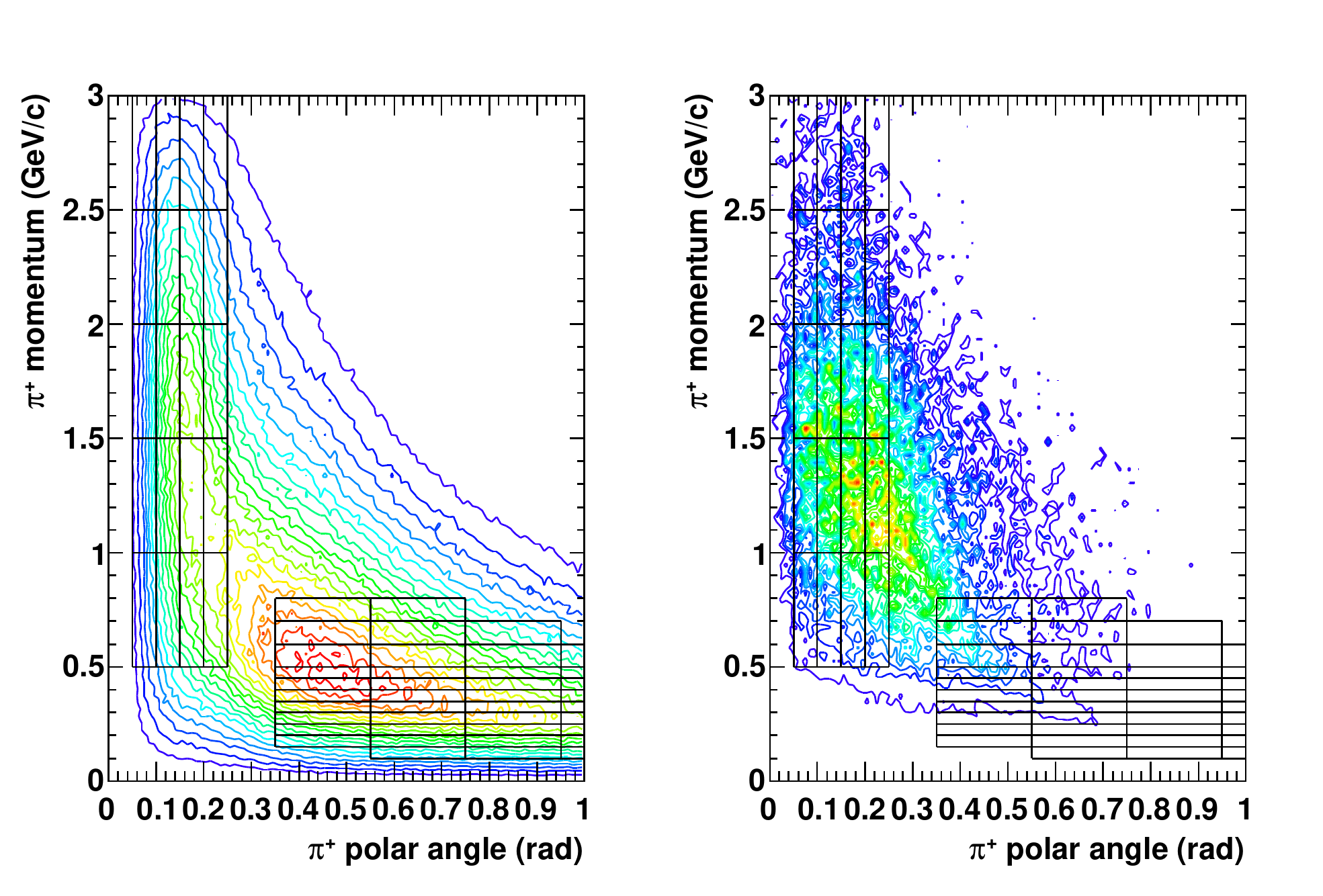}
}
  \caption{Polar angle versus momentum distributions for positive pions at target exit (left) and after reweighting with the 
  muon neutrino spectrum (right). The grids represent the binning used in the HARP analysis.}
 \label{fig:pionPHS}
\end{center}
\end{figure}
In Fig.~\ref{fig:pionPHS}
the $p$ versus $\theta$ distribution of positive pions at target exit is shown before (left) and after (right)
the above-mentioned reweighting.
While the inclusive sample peaks at $p \simeq$~0.5~GeV/c and $\theta \simeq$~0.5 rad, 
the pions which mainly contribute to the flux peak at higher momentum ($p \simeq$~1.2~GeV/c) and smaller angle ($\theta \simeq$ 0.2~rad). 
The grids superimposed to the distributions represent the binning used in the analysis of the HARP hadro-production experiment. 
The interesting phase space region is mainly covered by the so called ``small angle'' analysis for which only ``thin'' target\footnote{A cylinder with 1.95~cm length and 1.5~cm radius.} 
data exist at 5~GeV proton kinetic energy for carbon\cite{HARP}. 

We used these data to assess the size of the systematic effects introduced by the uncertainty on hadro-production. HARP pion differential cross sections have been compared with the expectations from GEANT4-QGSP and FLUKA.
In the comparison the geometry of the targets used in the HARP experiment has been 
modeled with the two simulation programs\footnote{
The formula which we used reflects the approach used in the HARP publications:
$\frac{\rm{d}^2\sigma}{\rm{d}p_i\rm{d}\theta_j} = \frac{1}{N_{pot}}\frac{A}{N_A\rho t} N_{i,j}$
where $N_{ij}$ are the number of events in the bin ($i$, $j$) at true level, $N_{pot}$ is the number of simulated protons interactions in the sample,
$\rho$ the density of the graphite target, $A$ the atomic number of carbon and $t$ the target length.}. 

\begin{figure*}
\begin{center}
\resizebox{0.9\textwidth}{!}{
\includegraphics{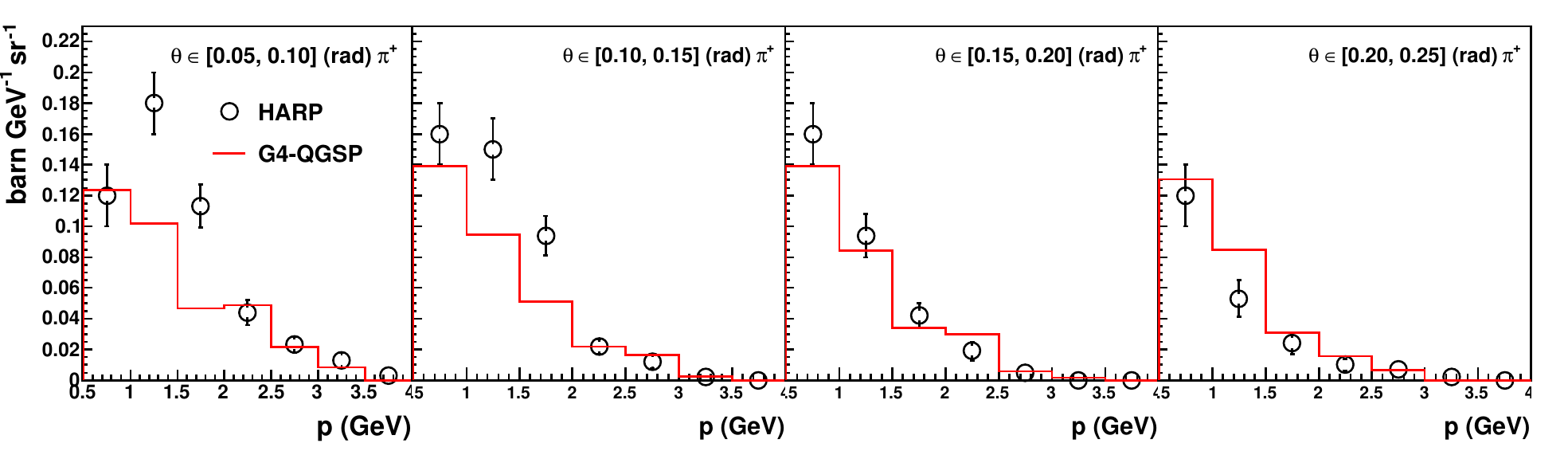}
}
\resizebox{0.9\textwidth}{!}{
\includegraphics{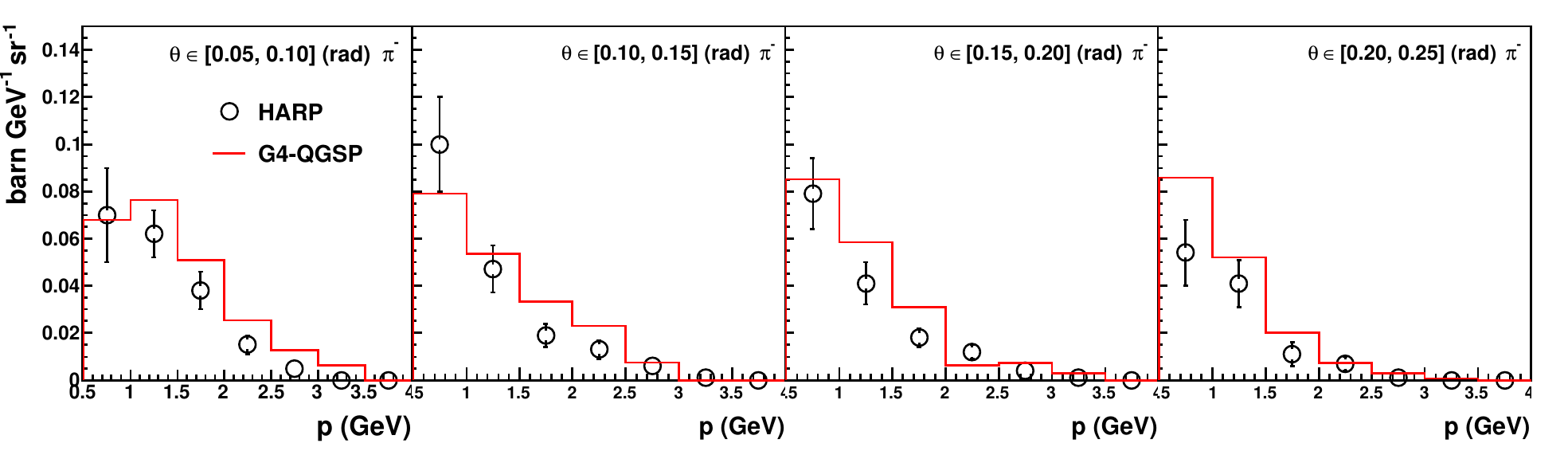} 
}
\caption{The HARP data points \cite{HARP} (bullets) are compared to the GEANT4-QGSP model (solid red histograms) 
for $\pi^+$ (upper row) and $\pi^-$ (lower row). Data correspond to 5~GeV protons impinging on a thin carbon target.}
 \label{fig:HARPcomp}
\end{center}
\end{figure*}

The comparison is shown in Fig.~\ref{fig:HARPcomp} for GEANT4-QGSP only.
The absolute normalization and the trend
of the experimental data is not too far from the simulation even though large discrepancies are observed in several bins.
Overall the QGSP model tends to be slightly closer to the data predicting softer momentum spectra.
As a systematic check we re-calculated the fluxes: 1) using the GEANT4-QGSP model in place of FLUKA; 
2) using FLUKA with reweighting factors derived from the comparison of FLUKA and HARP data.
The reweighting factors in the considered momentum and polar angle bins range between 0.14 and 1.6. The reweighted $\nu_\mu$ spectrum is peaked at lower energies where it exceeds the 
original one while the high energy tail is reduced.  
We will show the difference of these assumptions on the primary pion spectra at the level of sensitivities in the following section.
It must be noted that with this procedure we are neglecting possible effects related to the use of a long target and the effect of the 
slightly different proton kinetic energy.

\subsection{Discovery potential for $\theta_{13}\neq 0$ and CP violation.}

\begin{figure}
\centering
\resizebox{0.5\textwidth}{!}{%
\includegraphics{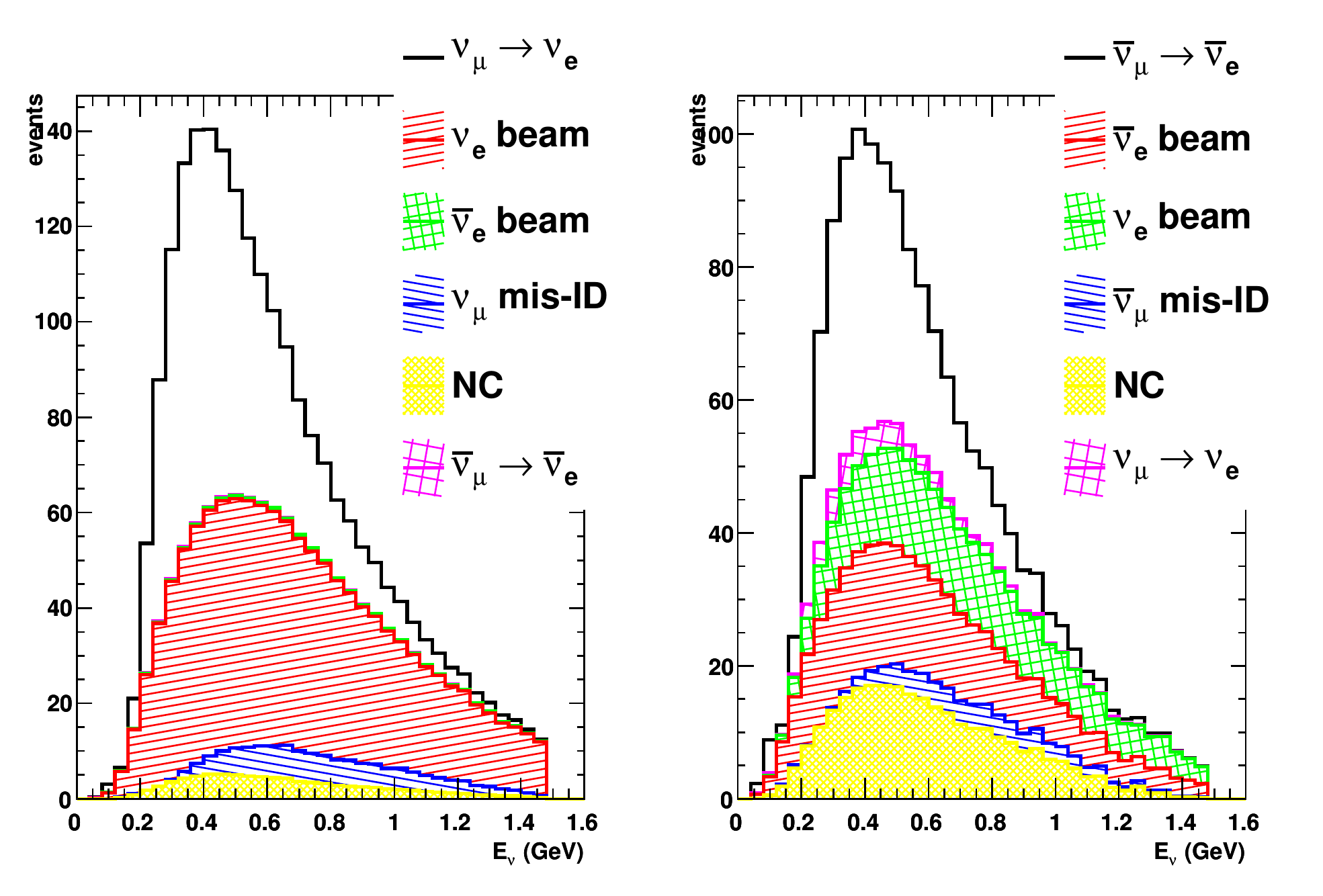}
}
\caption{Event rates in the MEMPHYS detector for $\sin^22\theta_{13}=0.01$ and $\delta_{CP}=0$
for 8+2 years of $\bar{\nu}$+$\nu$ running.}
\label{fig:eventrates}
\end{figure}

Event rates in the MEMPHYS detector 
are shown in Fig.~\ref{fig:eventrates} for $\sin^22\theta_{13}=0.01$ and $\delta_{CP}=0$. In positive focusing mode the main 
background comes from the intrinsic beam contamination (84\%) followed by muon events misidentified as electron events (9\%) and neutral current 
interactions (NC, 7\%). In the negative focusing mode the intrinsic beam contamination amounts to 65\% 
(35\% $\bar{\nu}_e^{CC}$ and 30\% $\nu_e^{CC}$), the misidentified event to 6\% and the NC interactions to 28\%.

\begin{figure}
\centering
\resizebox{0.5\textwidth}{!}{%
\includegraphics{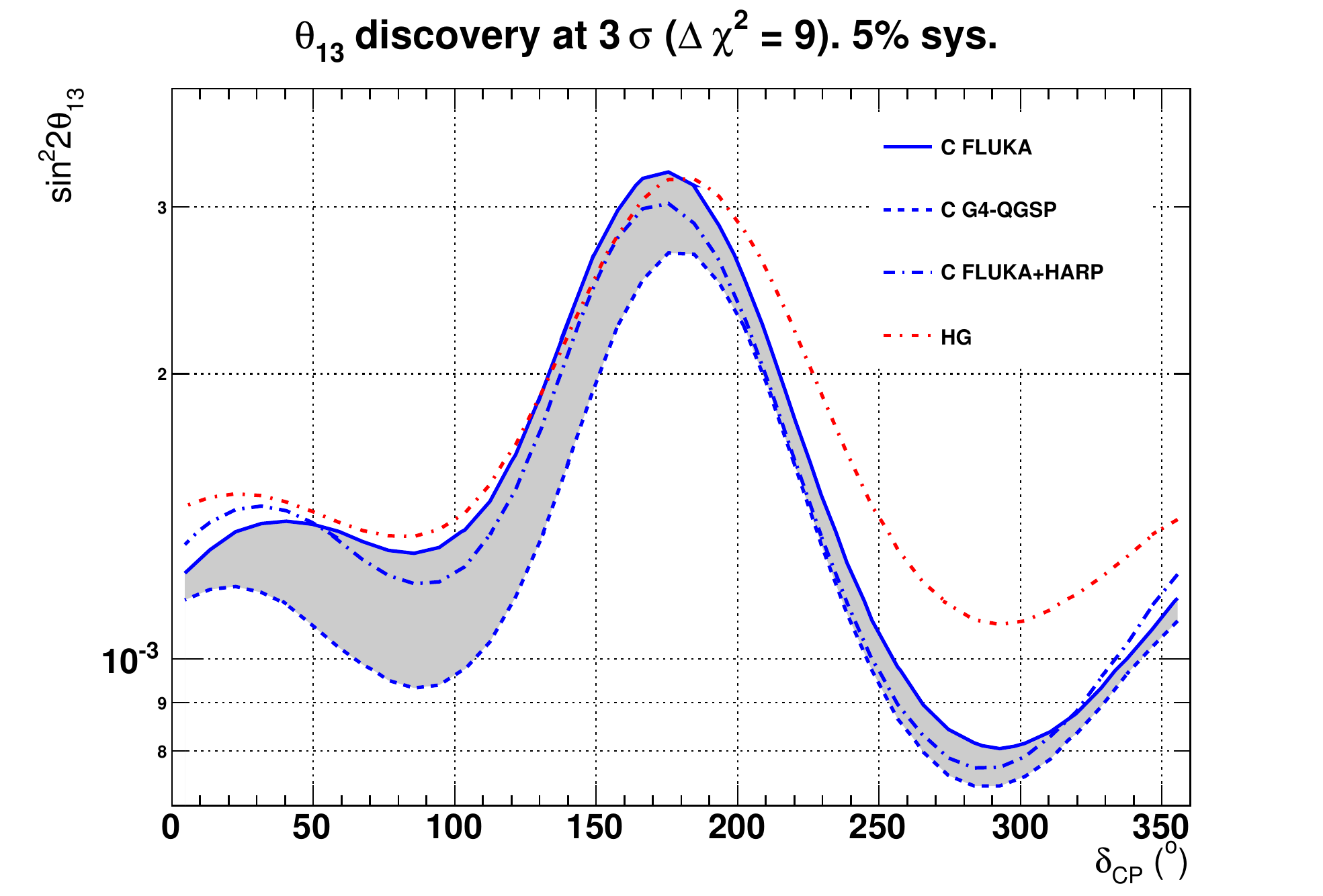}
}
\caption{$\theta_{13}$ discovery potential at 3~$\sigma$ level.  We set the ``true'' values of the parameters at $\Delta m^{2}_{12}=7\cdot 10^{-5}$ eV$^2$
$\Delta m^{2}_{31}=+2.43\cdot 10^{-3}$ eV$^2$ (normal hierarchy), $\theta_{12}=0.591$ and $\theta_{23}=\pi/4$. We included these parameters in the fit assuming a prior knowledge with an accuracy of 10\% for $\theta_{12}$, $\theta_{23}$,  5\% for $\Delta m^2_{31}$ and 3\% for at $\Delta m^2_{12}$ at 1 $\sigma$ level. The running time is (2$\nu$+8$\bar{\nu}$) years.
}
\label{fig:CPVa}
\end{figure}

In Fig.~\ref{fig:CPVa} the curves define the region in the space of true ($\sin^22\theta_{13}$, $\delta_{CP}$) 
for which a discovery of $\sin^22\theta_{13}\neq 0$ is possible at more than 3~$\sigma$. The limit as a function of
$\delta_{CP}$ is defined as the true value of  $\sin^22\theta_{13}$ for which a fit assuming $\theta_{13}=0$ gives
a $\Delta \chi^2=9$ as done, for example, in \cite{Campagne:2006yx}. In the fit a minimization on all the other oscillation parameters
is done and the two hypotheses for the mass hierarchy are taken into account.

The limit obtained with the previous setup associated with the mercury target is shown by the dash-dotted curve while the new limits are represented as a hatched band. 
The upper edge of the band (continuous line) refers to the FLUKA model of hadro-production, the lower edge (dotted) to the GEANT4-QGSP model, the one lying (mostly) in 
the middle (long dash-dotted) is obtained after reweighting FLUKA with the HARP data. 

\begin{figure}
\centering
\resizebox{0.5\textwidth}{!}{%
\includegraphics{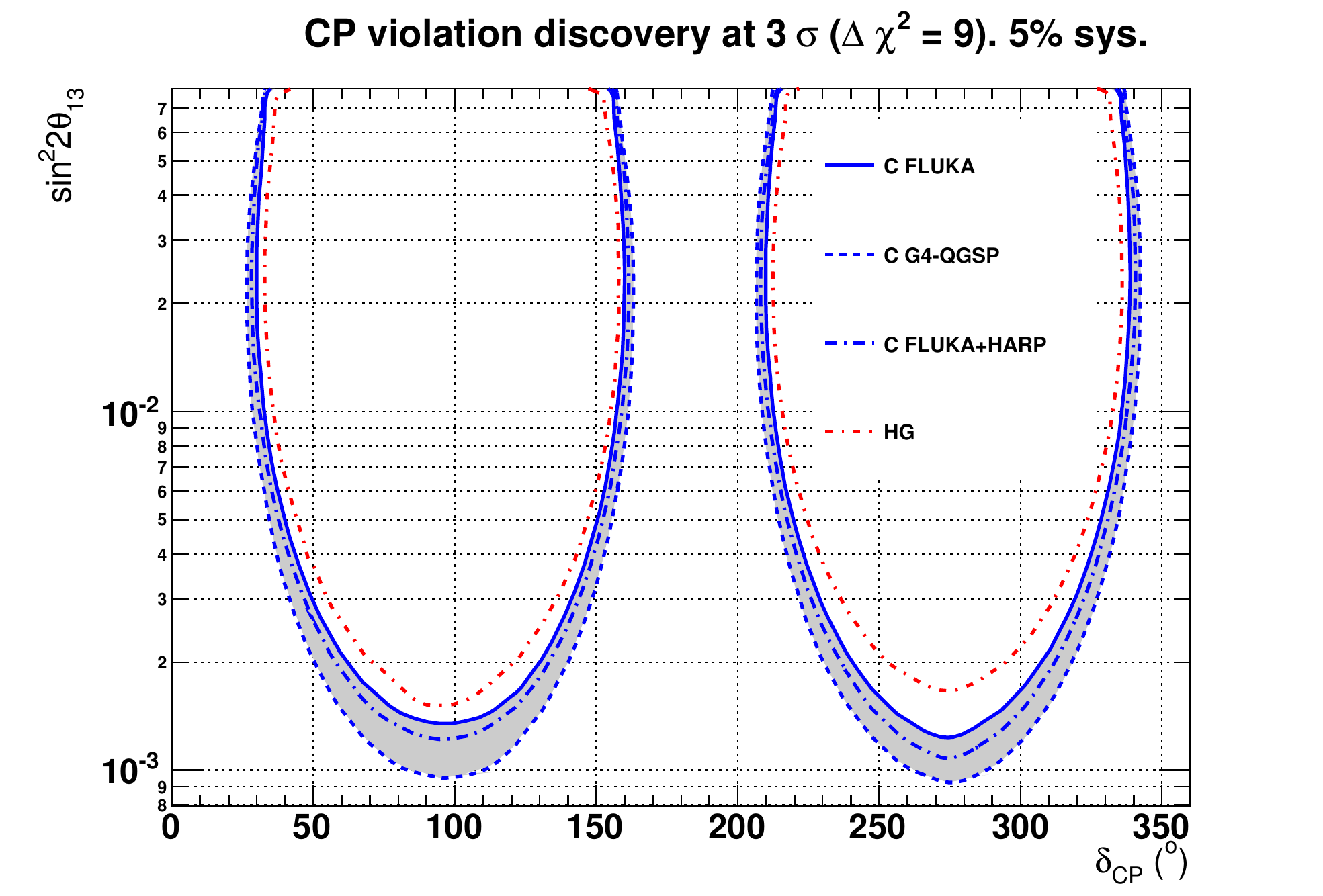}
}
\caption{CP violation discovery potential at 3~$\sigma$ level. See also Fig. \ref{fig:CPVa}.}
\label{fig:CPVb}
\end{figure}

The CPV discovery potential at the 3~$\sigma$ level is shown in Fig.~\ref{fig:CPVb}: discovery is possible in the region defined 
by the couple of curves. This means that in that region of the true ($\sin^22\theta_{13}$, $\delta_{CP}$) plane a fit done
under the CP conserving hypotheses ($\delta_{CP}=0, \pi$) gives for both choices
a $\Delta\chi^2>9$. The meaning of the symbols is the same as for the $\theta_{13}$ plot. 
The new limits generally improve those obtained with the previous design both for $\theta_{13}$ and CPV 
discovery. Considering the results obtained with FLUKA without reweighting (solid line) 
in the region dominated by the $\bar{\nu}$ run ($\delta < \pi$) the best limit is improved by about
4\% (from 1.35 to 1.30 10$^{-3}$) and in the complementary $\nu$-driven 
region by about 25\% (from 1.08 to 0.80 10$^{-3}$). The same criterion for the CPV discovery
gives and improvement of 10\% (from 1.50 to 1.35 10$^{-3}$) and 25\% (from 1.65 to 1.23 10$^{-3}$).
\section{Alternative target designs  \label{sect:alt}}
As undergoing studies in the context of the EUROnu design study
have shown technical challenges for the integrated horn-target system (IT), we
studied the performances of two additional configurations assuming:
\begin{itemize}
\item a graphite target separated from the horn (ST); 
\item a granular target composed of titanium spheres with diameters of $\mathcal{O}$(mm)
(PB, packed pebble bed target). 
\end{itemize}
while keeping the target geometry unchanged.

In the two cases we set for the inner radius of the horn ($R_1$) a value of 3~cm. The gap of 1.5~cm between the target 
and the horn is intended to accomodate the infrastructure needed for the target cooling.
As we showed earlier (right hand plot of Fig.~\ref{fig:Ltun}) the loss in performance due 
to an increase in the horn inner conductor radius can be compensated by an increase in the pulsed current. In the
following we will assume a value of 350~kA. This current is not far from the one used in T2K horns which amounts to 320~kA.
 
The granular target design was originally proposed in \cite{sievers}. Thanks to the favorable surface to volume
 ratio and the possibility to flow transversely the coolant within the interstices of the spheres (i.e. a high pressure flow of He gas), 
this target is expected to have a good behaviour even under the extreme irradiation conditions we are dealing with (1 MW beam power). 

The pion yields for the PB target have been calculated assuming a density reduction of
74\% which is obtainable using the optimal packing of the spheres (central packing).
We have verified that the reduction in density related to the granular structure tends to produce
a longitudinal profile for the emitted pions which resembles the one observed for the carbon target which
has been used in the optimization. This effect arises from the density reduction and the
shorter hadronic interaction length of Titanium with respect to graphite. 
Even if the achievable density reduction fraction would be lower than the assumed value
the inpact on the sensitivity would not be strong thanks to the fact that we are using a target which is relatively long
in units of $\lambda_I$.

\begin{figure}
\centering
\resizebox{0.5\textwidth}{!}{%
\includegraphics{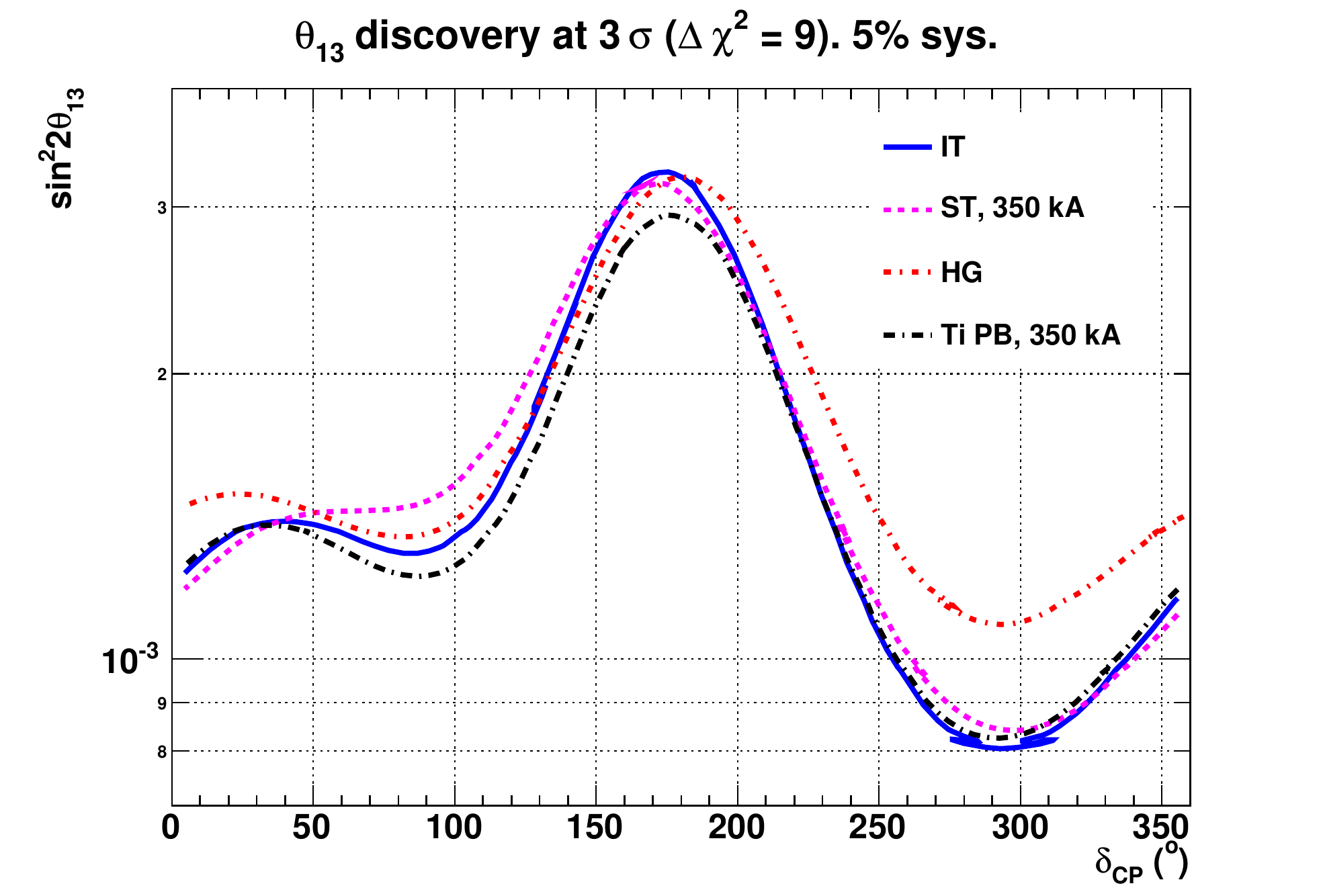}
}
\caption{$\theta_{13}$ discovery potential at 3~$\sigma$ level. See the text and Fig. \ref{fig:CPVa}.}
\label{fig:CPV1a}
\end{figure}

\begin{figure}
\centering
\resizebox{0.5\textwidth}{!}{%
\includegraphics{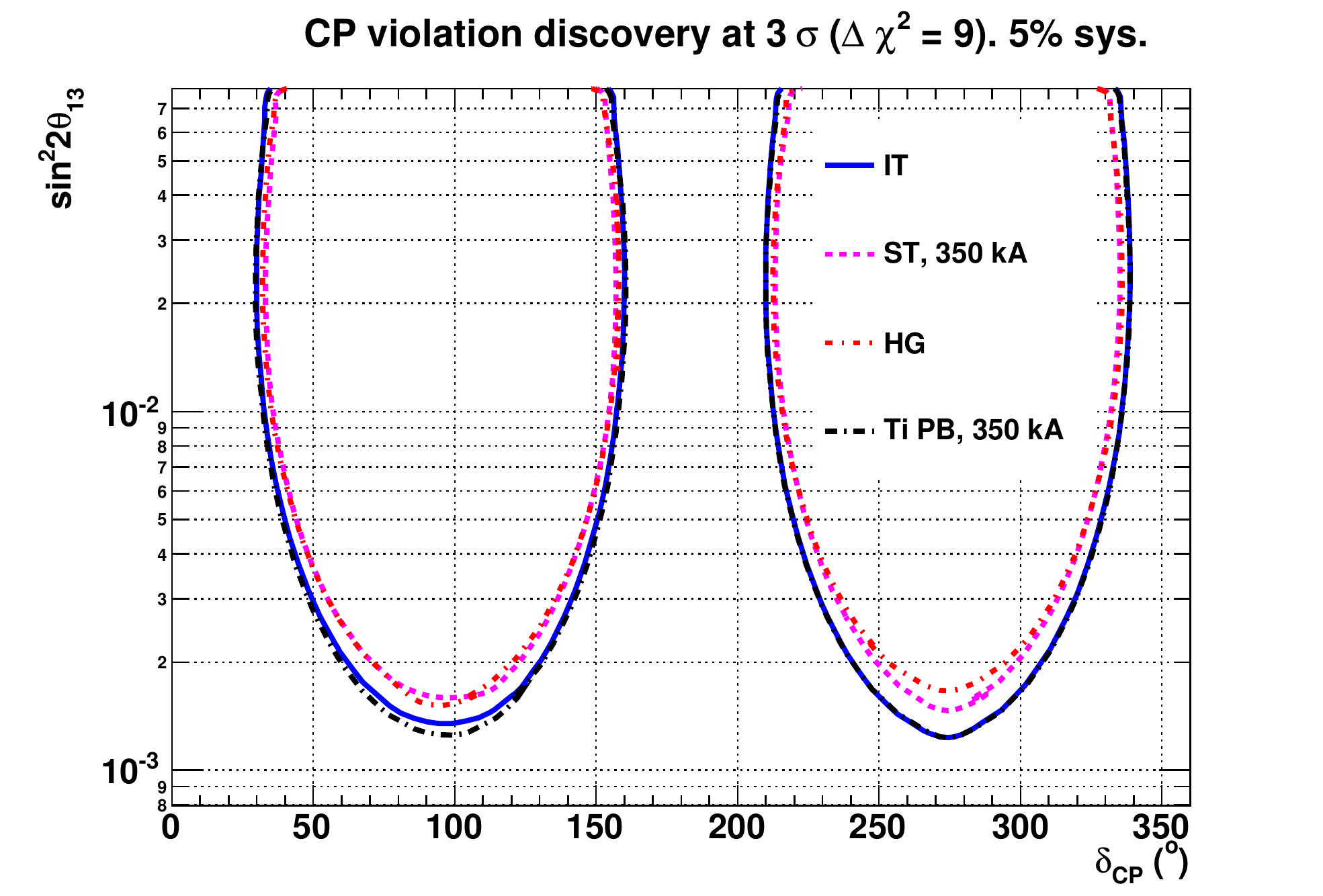}
}
\caption{CP violation discovery potential at 3~$\sigma$ level. See the text and Fig. \ref{fig:CPVa}.}
\label{fig:CPV1b}
\end{figure}

The discovery potentials for  $\theta_{13}\neq 0$
and CP violation for these two configurations (SB and PB) are compared to the ones obtained with the former design based on a mercury target (HG) and 
to the performance with the integrated target (IT) in Figs. \ref{fig:CPV1a} and \ref{fig:CPV1b}.  With respect to the IT design the solution with a separated monolithic graphite target and increased current (ST) introduces a moderate worsening of the limits; the PB solution gives practically unchanged performance for $\delta>\pi$ and some improvement for $\delta<\pi$. 
This effect can be understood in terms of a higher symmetry in the production of positive and negative pions for Titanium.

We can conclude that a PB target in association with the optimized horn represent possibly the most appealing solution in terms of
both physics performance and engineering.


\section{Summary  and conclusions
         \label{sec:summary}}

In this work the neutrino beam production for the SPL-Super Beam from CERN to
Fr\'ejus has been optimized for
the measurement
of the oscillation parameters $\theta_{13}$ and $\delta_{CP}$ aiming, at the same time, 
to a simplified setup in terms of technical feasibility.
 
The envisaged scenario adopts as a baseline the use of a solid target,
a single horn pulsed with a current of $300$~kA and a decay tunnel length of 25 m.
The target-horn system is foreseen to be replicated four times and operated in parallel.
With respect to the previous setup we abandoned the liquid mercury target option, 
the 600~kA pulsed reflector and shortened the decay tunnel (formerly 40 m).
The horn shape has been adapted to focus pions emerging from a long target through  
a systematic optimization procedure based on a forward--closed 
model  and a ranking
of the configurations relying on the sensitivity limit on $\theta_{13}$. 
The scenario involving four horns in parallel has been validated in terms of 
the achievable neutrino flux.
The final results of the sensitivities to $\sin^2 2 \theta_{13}$ and
$\delta_{CP}$ improve those obtained in \cite{Campagne:2004wt}\cite{Campagne:2006yx}.
The uncertainty on hadro-production has been addressed, for a graphite target, at the level of sensitivities 
by exploiting the data of the HARP experiment and different models (FLUKA and GEANT4-QGSP).

A strategy considering higher energy proton drivers and longer baselines is being addressed with a
similar simulation and optimization approach and will be the subject of a future publication.

I wish to thank M. Zito with whom I had a fruitful and intense collaboration, A. Cazes, J. E. Campagne, authors of the
original GEANT3 simulation, and C. Bobeth, M. Dracos, P. Cupial, C. Densham,
M. Mezzetto, M. Bonesini. I acknowledge the financial support of the European Community under the
European Commission Framework Programme 7 design study: EUROnu, project number
212372. The EC is not liable for any use that may be made of the information
contained herein.


\begin{thebibliography}{}
\bibitem{Ball:2000pj}
  A.~E.~Ball {\it et al.}
  N.I.M. A {\bf 451}, 359 (2000).
  N.I.M.  A {\bf 472}, 650 (2000),
  CERN-NUFACT-NOTE-42, CERN-OPEN-2000-339.
\bibitem{Gilardoni:2004kr}
S.~Gilardoni,
CERN-THESIS-2004-046, CERN-NUFACT-NOTE-141, GENEVA-THESE-3536, Jul. 2004.
\bibitem{Mezz03}
M. Mezzetto,
J.~Phys.~{\bf{G29}} (2003), 1781-1784, 
\bibitem{Gilardoni:2003us}
  S.~Gilardoni {\it et al.}
  J.\ Phys.\ G {\bf 29}, 1801 (2003).
\bibitem{Campagne:2004cd}
  J.E. Campagne,
  CERN-NUFACT-NOTE-138, Jul 2004.
\bibitem{Campagne:2004wt}
  J.E.~Campagne, A.~Cazes,
  Eur.\ Phys.\ J.\  C {\bf 45}, 643 (2006).
\bibitem{Campagne:2006yx}
  J.E.~Campagne {\it et al.},
  JHEP {\bf 0704}, 003 (2007).
\bibitem{cit:MERIT}
K.T. McDonald {\it et al.},
IPAC-2010-3527-3529.
\bibitem{Gilardoni:2001gv}
  S.~Gilardoni {\it et al.},
  CERN-OPEN-2001-042.
\bibitem{CERN:SPL:design}
  M.~Baylac {\em et al.},
  CERN-2006-006. 
  O.~Brunner {\em et al.},  
  CERN-AB-2008-067 BI/RF.
\bibitem{NuFact:acc-cmp:ring}
  M.~Aiba, 
  CERN-AB-2008-060 BI.
  CERN-AB-Note-2008-048 BI.
  E.~Benedetto, 
  CERN-BE-2009-037, CERN-NUFACT-NOTE-156, 
 AIP Conf. Proc. 1222 (2010) 283-287.
\bibitem{deBellefon:2006vq}
  A.~de Bellefon {\it et al.},
  hep-ex/0607026.
\bibitem{NUFACT09} A.~Longhin,
  AIP Conf.~Proc.~1222 (2010) 339-343.
\bibitem{ncampagne}
J.E. Campagne, CERN-NUFACT-NOTE-130, May 2003.
\bibitem{MINIBOONEbeam}A.A. Aguilar-Arevalo {\it et al.},
Phys. Rev. {\bf{D79}}, 072002 (2009).
\bibitem{Bobeth:2010:WP2:intTarget} 
  B.~Lepers {\it et al.},
  EUROnu-WP2-Note 10-01.
\bibitem{Nakadaira:2008zz}
  T.~Nakadaira {\it et al.},
  AIP Conf.\ Proc.\  {\bf 981}, 290 (2008).
\bibitem{Blondel:2000ph}
  A.~Blondel {\it et al.},
CERN-NUFACT-NOTE-53, CERN-OPEN-2000-343, Nov 2000.
CERN-NUFACT-NOTE-78, Sep 2001.
\bibitem{Cazes:1900zz}
  A.~Cazes,
  LAL-04-118, Ph.D. Thesis, Dec. 2004. 
\bibitem{Kformfact}
A. Lai {\it et al.} [NA48 Coll.]. 
Phys. Lett. {\bf{B647}}:341-350, 2007.
\bibitem{Battistoni:2007zzb}
  G.~Battistoni {\it et al.},
  AIP Conf.\ Proc.\  {\bf 896}, 31 (2007).
\bibitem{GEANT4}
N. I. M. A {\bf 506} (2003) 250-303.
\bibitem{Huber:2004ka}
  P.~Huber {\it et al.},
  Comput.\ Phys.\ Commun.\  {\bf 177}, 432 (2007).  
\bibitem{fluxesWEB} \verb+http://irfu.cea.fr/en/Phocea/Pisp/index.php?id=54+
\bibitem{HARP} 
M. G. Catanesi {\it et al.}, N.I.M. A {\bf 571} (2007) 527.
\mbox{M. Apollonio} {\it et al.},
Phys. Rev. {\bf{C80}} (2009) 035208. 
\bibitem{sievers} P. Pugnat, P. Sievers, CERN-NUFACT-NOTE-127. Dec. 2002.
\end{thebibliography}
\end{document}